\documentclass[aps, superscriptaddress, nofootinbib, preprintnumbers]{revtex4}

\usepackage{amsmath}	
\usepackage{amsthm}	
\usepackage{amssymb}	
\usepackage{datetime}
\usepackage{graphicx}
\usepackage{color}
\usepackage{verbatim}
\usepackage{bm}

\usepackage{hyperref}

\definecolor{grey}{rgb}{0.4,0.4,0.4}
\definecolor{dullmagenta}{rgb}{0.4,0,0.4}
\definecolor{darkblue}{rgb}{0,0,0.4}
\definecolor{midblue}{rgb}{0,0,0.5}
\definecolor{midred}{rgb}{0.5,0,0}
\definecolor{orange}{rgb}{1,0.5,0}
\definecolor{lightbrown}{rgb}{0.75,0.5,0.25}
\definecolor{tan}{cmyk}{0.14,0.42,0.56,0}
\definecolor{djunglegreen}{cmyk}{0.99,0,0.52,0}
\definecolor{lightgreen}{rgb}{0,1,0}
\definecolor{olivegreen}{cmyk}{0.64,0,0.95,0.40}
\definecolor{midgreen}{rgb}{0.0,0.675,0.0}
\definecolor{darkgreen}{rgb}{0,0.5,0}

\newcommand{\vs}{\vspace}

\renewcommand{\.}{\hspace{0.5mm}}

\newcommand{\Rrm}{\ensuremath{\mathrm{R}}}

\newcommand{\crm}{\ensuremath{\mathrm{c}}}
\newcommand{\drm}{\ensuremath{\mathrm{d}}}
\newcommand{\erm}{\ensuremath{\mathrm{e}}}
\newcommand{\frm}{\ensuremath{\mathrm{f}}}
\newcommand{\grm}{\ensuremath{\mathrm{g}}}

\newcommand{\irm}{\ensuremath{\mathrm{i}}}

\newcommand{\srm}{\ensuremath{\mathrm{s}}}

\newcommand{\Ecal}{\ensuremath{\mathcal{E}}}

\newcommand{\Ocal}{\ensuremath{\mathcal{O}}}
\newcommand{\Pcal}{\ensuremath{\mathcal{P}}}

\renewcommand{\d}{\ensuremath{\mathrm{d}}}

\newcommand{\eg}{eg.}
\newcommand{\ie}{ie.}

\newcommand{\rhs}{r.h.s.}

\newcommand{\cf}{cf.}

\let\baraccent=\= 
\renewcommand{\=}[1]{\stackrel{#1}{=}} 

\theoremstyle{definition}

\theoremstyle{remark}

\smallskipamount

\settimeformat{ampmtime}

\usepackage{float}

\begin{document}

\title{PRIMORDIAL BLACK HOLES AS DARK MATTER}

\author{Bernard Carr}
\email{b.j.carr@qmul.ac.uk}
\affiliation{Department of Physics and Astronomy,
	Queen Mary University of London,
	Mile End Road,
	London E1 4NS,
	United Kingdom}

\author{Florian K{\"u}hnel}
\email{florian.kuhnel@fysik.su.se}
\affiliation{The Oskar Klein Centre for Cosmoparticle Physics,
	Department of Physics,
	Stockholm University,
	AlbaNova,
	SE--106\.91 Stockholm,
	Sweden}

\author{Marit Sandstad}
\email{marit.sandstad@astro.uio.no}
\affiliation{Nordita,
	KTH Royal Institute of Technology and Stockholm University,
	Roslagstullsbacken 23,
	SE--106\.91 Stockholm,
	Sweden}

\date{\formatdate{\day}{\month}{\year}, \currenttime}

\begin{abstract}
The possibility that the dark matter comprises primordial black holes (PBHs) is considered, with particular emphasis on the currently allowed mass windows at $10^{16}$ -- $10^{17}\,$g, $10^{20}$ -- $10^{24}\,$g and $1$ -- $10^{3}\,M_{\odot}$. The Planck mass relics of smaller evaporating PBHs are also considered. All relevant constraints (lensing, dynamical, large-scale structure and accretion) are reviewed and various effects necessary for a precise calculation of the PBH abundance (non-Gaussianity, non-sphericity, critical collapse and merging) are accounted for. It is difficult to put all the dark matter in PBHs if their mass function is monochromatic but this is still possible if the mass function is extended, as expected in many scenarios. A novel procedure for confronting observational constraints with an extended PBH mass spectrum is therefore introduced. This applies for arbitrary constraints and a wide range of PBH formation models, and allows us to identify which model-independent conclusions can be drawn from constraints over all mass ranges. We focus particularly on PBHs generated by inflation, pointing out which effects in the formation process influence the mapping from the inflationary power spectrum to the PBH mass function. We then apply our scheme to two specific inflationary models in which PBHs provide the dark matter. The possibility that the dark matter is in intermediate-mass PBHs of $1$ -- $10^{3}\,M_{\odot}$ is of special interest in view of the recent detection of black-hole mergers by LIGO. The possibility of Planck relics is also intriguing but virtually untestable.
\end{abstract}

\preprint{NORDITA-2016-83}

\maketitle

\section{Introduction}
\label{sec:Introduction}

\noindent
Primordial black holes (PBHs) have been a source of intense interest for nearly 50 years \cite{ZeldovichNovikov69}, despite the fact that there is still no evidence for them. One reason for this interest is that only PBHs could be small enough for Hawking radiation to be important \cite{Hawking:1974rv}. This has not yet been confirmed experimentally and there remain major conceptual puzzles associated with the process, with Hawking himself still grappling with these \cite{Hawking:2016msc}. Nevertheless, this discovery is generally recognised as one of the key developments in 20th century physics because it beautifully unifies general relativity, quantum mechanics and thermodynamics. The fact that Hawking was only led to this discovery through contemplating the properties of PBHs illustrates that it can be useful to study something even if it may not exist!

PBHs smaller than about $10^{15}\,\grm$ would have evaporated by now with many interesting cosmological consequences. Studies of such consequences have placed useful constraints on models of the early Universe and, more positively, evaporating PBHs have been invoked to explain certain features: for example, the extragalactic \cite{Page:1976wx} and Galactic \cite{Lehoucq:2009ge} $\gamma$-ray backgrounds, antimatter in cosmic rays \cite{Barrau:1999sk}, the annihilation line radiation from the Galactic centre \cite{Bambi:2008kx}, the reionisation of the pregalactic medium \cite{Belotsky:2014twa} and some short-period $\gamma$-ray bursts \cite{Cline:1996zg}. For more comprehensive references, see recent articles by Khlopov \cite{Khlopov:2008qy} and Carr \textit{et al.} \cite{Carr:2009jm} and the book by Calmet \textit{et al.} \cite{Calmet:2014dea}. However, there are usually other possible explanations for these features, so there is no definitive evidence for evaporating PBHs.

Attention has therefore shifted to the PBHs larger than $10^{15}\,\grm$, which are unaffected by Hawking radiation. Such PBHs might have various astrophysical consequences, such as providing seeds for the supermassive black holes in galactic nuclei \cite{Bean:2002kx}, the generation of large-scale structure through Poisson fluctuations \cite{Afshordi:2003zb} and important effects on the thermal and ionisation history of the Universe \cite{Ricotti:2007au}. For a recent review, in which a particular PBH-producing model is shown to solve these and several other observational problems, see Ref.~\cite{Dolgov:2016qsm}. But perhaps the most exciting possibility{\,---\,}and the main focus of this paper{\,---\,}is that they could provide the dark matter which comprises $25\%$ of the critical density, an idea that goes back to the earliest days of PBH research \cite{1975Natur.253..251C}. Since PBHs formed in the radiation-dominated era, they are not subject to the well-known big-bang nucleosynthesis (BBNS) constraint that baryons can have at most $5\%$ of the critical density \cite{Cyburt:2003fe}. They should therefore be classed as non-baryonic and from a dynamical perspective they behave like any other form of cold dark matter (CDM).

There is still no compelling evidence that PBHs provide the dark matter, but nor is there for any of the more traditional CDM candidates. One favored candidate is a Weakly Interacting Massive Particle (WIMP), such as the lightest supersymmetric particle \cite{Jungman:1995df} or the axion \cite{Preskill:1982cy}, but 30 years of accelerator experiments and direct dark-matter searches have not confirmed the existence of these particles \cite{DiValentino:2014zna}. One should not be too deterred by this{\,---\,}after all, the existence of gravitational waves was predicted 100 years ago, the first searches began nearly 50 years ago \cite{Weber:1969bz} and they were only finally detected by LIGO a few months ago \cite{Abbott:2016blz}. Nevertheless, even some theorists have become pessimistic about WIMPs \cite{Frampton:2015xza}, so this does encourage the search for alternative candidates.

There was a flurry of excitement around the PBH dark-matter hypothesis in the 1990s, when the Massive Astrophysical Compact Halo Object (MACHO) microlensing results \cite{Alcock:1996yv} suggested that the dark matter could be in compact objects of mass $0.5\,M_{\odot}$ since alternative MACHO candidates could be excluded and PBHs of this mass might naturally form at the quark-hadron phase transition at $10^{-5}\,\srm$ \cite{Jedamzik:1998hc}. Subsequently, however, it was shown that such objects could comprise only $20\%$ of the dark matter and indeed the entire mass range $10^{-7}\,M_{\odot}$ to $10\,M_{\odot}$ was excluded from providing the dark matter \cite{Tisserand:2006zx}. At one point there were claims to have discovered a critical density of $10^{-3}\,M_{\odot}$ PBHs through the microlensing of quasars \cite{1993Natur.366..242H} but this claim was met with scepticism \cite{Zackrisson:2003wu} and would seem to be incompatible with other lensing constraints. Also femtolensing of $\gamma$-ray bursts excluded $10^{17}$ -- $10^{20}\,\grm$ PBHs \cite{Nemiroff:2001bp}, microlensing of quasars constrained $10^{-3}$ -- $60\,M_{\odot}$ PBHs \cite{1994ApJ...424..550D} and millilensing of compact radio sources excluded $10^{6}$ -- $10^{9}\,M_{\odot}$ PBHs \cite{Wilkinson:2001vv} from explaining the dark matter. Dynamical constraints associated with the tidal disruption of globular clusters, the heating of the Galactic disc and the dragging of halo objects into the Galactic nucleus by dynamical friction excluded PBHs in the mass range above $10^{5}\,M_{\odot}$ \cite{Carr:1997cn}.

About a decade ago, these lensing and dynamical constraints appeared to allow three mass ranges in which PBHs could provide the dark matter \cite{Barrau:2003xp}: the subatomic-size range ($10^{16}$ -- $10^{17}\,\grm$), the sublunar mass range ($10^{20}$ -- $10^{26}\,\grm$) and what is sometimes termed the intermediate-mass black hole (IMBH) range ($10$ -- $10^{5}\,M_{\odot}$).\footnote{This term is commonly used to describe black holes intermediate between those which derive from the collapse of ordinary stars and the supermassive ones which derive from general relativistic instability, perhaps the remnants of a first generation of Population III stars larger than $10^{2}\,M_{\odot}$. Here we use it in a more extended sense to include the $\Ocal(10)\,M_{\odot}$ black holes detected by LIGO.} The lowest range may now be excluded by Galactic $\gamma$-ray observations \cite{Carr:2016hva} and the middle range{\,---\,}although the first to be proposed as a PBH dark-matter candidate \cite{1975Natur.253..251C}{\,---\,}is under tension because such PBHs would be captured by stars, whose neutron star or white-dwarf remnants would subsequently be destroyed by accretion \cite{Capela:2013yf}. One problem with PBHs in the IMBH range is that such objects would disrupt wide binaries in the Galactic disc. It was originally claimed that this would exclude objects above $400\,M_{\odot}$ \cite{Quinn:2009zg} but more recent studies may reduce this mass \cite{Monroy-Rodriguez:2014ula}, so the narrow window between the microlensing and wide-binary bounds is shrinking. Nevertheless, this suggestion is topical because PBHs in the IMBH range could naturally arise in the inflationary scenario \cite{Frampton:2010sw} and might also explain the sort of massive black-hole mergers observed by LIGO \cite{Bird:2016dcv}. The suggestion that LIGO could detect gravitational waves from a population of IMBHs comprising the dark matter was originally proposed in the context of the Population III ``VMO'' scenario by Bond \& Carr \cite{1984MNRAS.207..585B}. This is now regarded as unlikely, since the precursor stars would be baryonic and therefore subject to the BBNS constraint, but the same possibility applies for IMBHs of primordial origin.

Most of the PBH dark-matter proposals assume that the mass function of the black holes is very narrow (\ie~nearly monochromatic). However, this is unrealistic and in most scenarios one would expect the mass function to be extended. In particular, this arises if they form with the low mass tail expected in critical collapse \cite{Niemeyer:1997mt}. Indeed, it has been claimed that this would allow PBHs somewhat above $10^{15}\,\grm$ to contribute to both the dark matter and the $\gamma$-ray background \cite{Yokoyama:1998xd}. However, this assumes that the ``bare'' PBH mass function (\ie~without the low-mass tail) has a monochromatic form and recently it has been realised that the tail could have a wider variety of forms if one drops this assumption \cite{Kuhnel:2015vtw}. There are also many scenarios (\eg~PBH formation from the collapse of cosmic strings) in which even the bare mass function may be extended.

This raises two interesting questions: (1) Is there still a mass window in which PBHs could provide all of the dark matter without violating the bounds in other mass ranges? (2) If there is no mass scale at which PBHs could provide all the dark matter for a nearly monochromatic mass function, could they still provide it by being spread out in mass? In this paper we will show how to address these questions for both a specific extended mass function and for the more general situation. As far as we are aware, this issue has not been discussed in the literature before and we will apply this methodology to the three mass ranges mentioned above. There are subtleties involved when applying differential limits to models with extended mass distributions, especially when the experimental bounds come without a mention of the bin size or when different limits using different bin sizes are combined. In order to make model-independent statements, we also discuss which physical effects need to be taken into account in confronting a model capable of yielding a significant PBH abundance with relevant constraints. This includes critical collapse, non-sphericity and non-Gaussianity, all of which we investigate quantitatively for two specific inflationary models. We also discuss qualitatively some other extensions of the standard model. In principle, this approach could constrain the primordial curvature perturbations even if PBHs are excluded as dark-matter candidates \cite{Josan:2009qn}.

The plan of this paper is as follows: In Sec.~\ref{sec:ModelsInRelevantRanges} we review the PBH formation mechanisms. In Sec.~\ref{sec:SpecificModelsInRelevantRanges} we give a more detailed description of two inflationary models for PBH formation, later used to demonstrate our methodology. In Sec.~\ref{sec:RealisticMassFunctions} we consider some issues which are important in going from the initial curvature or density power spectrum to the PBH mass function. Many of these issues are not fully understood, but they may have a large impact on the final mass function, so their proper treatment is crucial in drawing conclusions about inflationary models from PBH constraints. In Sec.~\ref{sec:Constraints} we review the constraints for PBHs in the non-evaporating mass ranges above $10^{15}$g, concentrating particularly on the weakly constrained region around $\Ocal(10)\,M_{\odot}$. In Sec.~\ref{sec:ExtendedPBH} we explore how an extended mass function can still contain all the dark matter. In order to make model-independent exclusions, we develop a methodology for applying arbitrary constraints to any form of extended mass function. In Sec.~\ref{sec:LIGO} we discuss the new opportunities offered by gravitational-wave astronomy and the possible implications of the LIGO events. In Sec.~\ref{sec:Summary-and-Outlook} we summarise our results and outline their implications for future PBH searches.

\section{Introduction to PBH formation}
\label{sec:ModelsInRelevantRanges}

\noindent
PBHs could have been produced during the early Universe due to various mechanisms. For all of these, the increased cosmological energy density at early times plays a major role \cite{Hawking:1971ei, Carr:1974nx}, yielding a rough connection between the PBH mass and the horizon mass at formation:
\begin{align}
	M
		&\sim
								\frac{ c^{3}\, t }{ G }
		\sim
								10^{15}\,
								\left(
									\frac{t}{10^{-23}\,\srm}
								\right)\,
								\grm
								\, .
								\label{eq:Moft}
\end{align}
Hence PBHs could span an enormous mass range: those formed at the Planck time ($10^{-43}\,\srm$) would have the Planck mass ($10^{-5}\,\grm$), whereas those formed at $1\,\srm$ would be as large as $10^{5}\, M_{\odot}$, comparable to the mass of the holes thought to reside in galactic nuclei. By contrast, black holes forming at the present epoch (eg. in the final stages of stellar evolution) could never be smaller than about $1\,M_{\odot}$. In some circumstances PBHs may form over an extended period, corresponding to a wide range of masses. Even if they form at a single epoch, their mass mass spectrum could still extend much below the horizon mass due to ``critical phenomena'' \cite{Gundlach:1999cu, Gundlach:2002sx, Niemeyer:1997mt, Niemeyer:1999ak, Shibata:1999zs, Musco:2004ak, Musco:2008hv, Musco:2012au, Kuhnel:2015vtw}, although most of the PBH density would still be in the most massive ones. We return to these points in Sec.~\ref{sec:RealisticMassFunctions}.

\subsection{Formation Mechanisms}
\label{sec:Formation-mechanisms}

The high density of the early Universe is a necessary but not sufficient condition for PBH formation. One possibility is that there were large primordial inhomogeneities, so that overdense regions could stop expanding and recollapse. In this context, Eq.~\eqref{eq:Moft} can be replaced by the more precise relationship \cite{Carr:2009jm}
\begin{equation}
	M
		=
								\gamma\,M_\mathrm{PH}
		\approx
								2.03 \times 10^{5}\,
								\gamma
								\left( \frac{t}{1\,\srm} \right)
								M_{\odot}
								\, .
								\label{mass}
\end{equation}
Here $\gamma$ is a numerical factor which depends on the details of gravitational collapse. A simple analytical calculation suggests that it is around $(1 / \sqrt{3\,})^{3} \approx 0.2$ during the radiation era \cite{Carr:1975qj}, although the first hydrodynamical calculations gave a somewhat smaller value \cite{1978SvA....22..129N}. The favoured value has subsequently fluctuated as people have performed more sophisticated computations but now seems to have settled at a value of around $0.4$ \cite{Green:2004wb}.

It has been claimed that a PBH cannot be much larger than the value given by Eq.~\eqref{eq:Moft} at formation, else it would be a separate closed Universe rather than a part of our Universe \cite{Carr:1974nx, Harada:2004pe}. While there is a separate-Universe scale and Eq.~\eqref{eq:Moft} does indeed give an upper limit on the PBH mass, the original argument is not correct because the PBH mass necessarily goes to zero on the separate-Universe scale \cite{Kopp:2010sh, Carr:2014pga}. However, the effective value of $\gamma$ in Eq.~\eqref{mass} could exceed $1$ in some circumstances. In particular, if a PBH grows as a result of accretion, its {\it final} mass could well be larger than the horizon mass at formation.

As discussed in numerous papers, the quantum fluctuations arising in various inflationary scenarios are a possible source of PBHs. In some of these scenarios the fluctuations generated by inflation are ``blue'' (i.e.~decrease with increasing scale) and this means that the PBHs form shortly after reheating \cite{Carr:1993aq, Carr:1994ar, Leach:2000ea, Kohri:2007gq}. Others involve some form of ``designer'' inflation, in which the power spectrum of the fluctuations{\,---\,}and hence PBH production{\,---\,}peaks on some scale \cite{Hodges:1990bf, Ivanov:1994pa, Yokoyama:1995ex, Yokoyama:1998pt, Yokoyama:1998qw, Kawasaki:1998vx, Yokoyama:1999xi, Easther:1999ws, Kanazawa:2000ea, Blais:2002gw, Blais:2002nd, Barrau:2002ru, Chongchitnan:2006wx, Nozari:2007kv, Saito:2008em, Lyth:2001nq, Lyth:2002my, Kasuya:2009up, Kohri:2012yw, Kawasaki:2012wr, Josan:2010vn, Belotsky:2014kca, Clesse:2010iz, Kodama:2011vs, Bugaev:2008gw, Cheng:2016qzb}. In other scenarios, the fluctuations have a ``running index'', so that the amplitude increases on smaller scales but not according to a simple power law \cite{GarciaBellido:1996qt, Randall:1995dj, Stewart:1996ey, Stewart:1997wg, Lidsey:2001nj, Easther:2004qs, Lyth:2005ze, Kohri:2007qn, Bugaev:2008bi, Alabidi:2009bk, Leach:2000ea, Drees:2011yz, Drees:2011hb, Drees:2012sz, Kuhnel:2015vtw}. PBH formation may also occur due to some sort of parametric resonance effect before reheating \cite{Taruya:1998cz, Bassett:2000ha, Green:2000he, Finelli:2000gi, Kawaguchi:2007fz, Kawasaki:2007zz, Frampton:2010sw}. In this case, the fluctuations tend to peak on a scale associated with reheating. This is usually very small but several scenarios involve a secondary inflationary phase which boosts this scale into the macroscopic domain. Recently there has been a lot of interest in the formation of intermediate-mass PBHs in the ``waterfall'' scenario \cite{Frampton:2010sw, Bugaev:2011qt, Clesse:2015wea, Kawasaki:2015ppx} and the generation of PBH dark matter in supergravity inflation models is discussed in Ref.~\cite{Kawasaki:2016pql}. It has been claimed \cite{Young:2015kda} that any multiple-field inflationary model which generates enough PBHs to explain the dark matter is ruled out because it also generates an unacceptably large isocurvature perturbation due to the inherent non-Gaussianities in these models. We will discuss this in more detail in Sec.~\ref{sec:RealisticMassFunctions}.

Whatever the source of the inhomogeneities, PBH formation would be enhanced if there was a sudden reduction in the pressure{\,---\,}for example, at the QCD era \cite{Jedamzik:1996mr, Widerin:1998my, Jedamzik:1999am}{\,---\,}or if the early Universe went through a dustlike phase at early times as a result of either being dominated by non-relativistic particles for a period \cite{Khlopov:1980mg, 1981SvA....25..406P, 1982SvA....26..391P} or undergoing slow reheating after inflation \cite{Khlopov:1985jw, Carr:1994ar}. Another possibility is that PBHs might have formed spontaneously at some sort of phase transition, even if there were no prior inhomogeneities, for example from bubble collisions \cite{Crawford:1982yz, Hawking:1982ga, Kodama:1982sf, La:1989st, Moss:1994iq, 1998AstL...24..413K, Konoplich:1999qq} or from the collapse of cosmic strings \cite{Hogan:1984zb, Hawking:1987bn, Polnarev:1988dh, Garriga:1993gj, Caldwell:1995fu, Cheng:1996du, MacGibbon:1997pu, Hansen:1999su, Nagasawa:2005hv}, necklaces \cite{Matsuda:2005ez, Lake:2009nq} or domain walls \cite{Berezin:1982ur, Caldwell:1996pt, Khlopov:2000js, Rubin:2000dq, Rubin:2001yw, Dokuchaev:2004kr}. Braneworld scenarios with a modified-gravity scale of $\sim 1\,{\rm TeV}$ may lead to the production of lunar-mass PBHs \cite{Inoue:2003di}.

\subsection{Collapse Fraction}
\label{sec:Collapse-fraction}

The fraction of the mass of the Universe in PBHs on some mass-scale $M$ is epoch-dependent but its value at the formation epoch of the PBHs is denoted by $\beta( M )$.The current density parameter $\Omega_{\mathrm{PBH}}$ (in units of the critical density) associated with unevaporated PBHs which form at a redshift $z$ or time $t$ is roughly related to $\beta$ by \cite{Carr:1975qj}
\begin{equation}
	\Omega_{\mathrm{PBH}}
		\simeq
								\beta\,\Omega_\mathrm r\,( 1 + z )
		\sim
								10^{6}\,\beta \left( \frac{t}{1\,\srm} \right)^{\!-1/2}
		\sim
								10^{18}\,\beta \left( \frac{M}{10^{15}\,\grm} \right)^{\!-1/2}
		\quad
								( M > 10^{15}\,\grm )
								\, ,
								\label{eq:roughomega}
\end{equation}
where $ \Omega_{\mathrm r} \sim 10^{-4}$ is the density parameter of the cosmic microwave background (CMB) and we have used Eq.~\eqref{eq:Moft}. The $(1 + z)$ factor arises because the radiation density scales as $(1 + z)^{4}$\,, whereas the PBH density scales as $(1 + z)^{3}$\,. Any limit on $\Omega_{\mathrm{PBH}}$ therefore places a constraint on $\beta( M )$\,. The parameter $\Omega_{\mathrm{PBH}}$ must be interpreted with care for PBHs which have already evaporated, since they no longer contribute to the cosmological density. Note that Eq.~\eqref{eq:roughomega} assumes that the PBHs form in the radiation-dominated era, in which case $\beta$ is necessarily small.

We can determine the relationship \eqref{eq:roughomega} more precisely for the standard $\Lambda$CDM model, in which the age of the Universe is $t_{0} = 13.8\,\mathrm{Gyr}$, the Hubble parameter is $h = 0.68$ \cite{Ade:2015lrj} and the time of photon decoupling is $t_{\mathrm{dec}} = 380\,\mathrm{kyr}$ \cite{Hinshaw:2008kr}. If the PBHs have a monochromatic mass function, then the fraction of the Universe's mass in PBHs at their formation time $t_{\irm}$ is related to their number density at $t_{\irm}$ and $t_{0}$ by \cite{Carr:2009jm}
\begin{equation}
	\beta( M )
		\equiv
								\frac{M\,n_{\mathrm{PBH}}(t_{\irm})}{\rho(t_{\irm})}
		\approx
								7.98 \times 10^{-29}\,
								\gamma^{-1/2}\,
								\left( \frac{g_{* \irm}}{106.75} \right)^{\!1/4}\,
								\left( \frac{M}{M_{\odot}} \right)^{\!3/2}\,
								\left( \frac{n_{\mathrm{PBH}}(t_{0})}{1\,\mathrm{Gpc}^{-3}} \right)
								,
								\label{eq:beta}
\end{equation}
where we have used Eq.~\eqref{mass} and $g_{* \irm}$ is the number of relativistic degrees of freedom at PBH formation. $g_{* \irm}$ is normalised to its value at around $10^{-5}\,\srm$ since it does not increase much before that in the Standard Model and that is the period in which most PBHs are likely to form. The current density parameter for PBHs which have not yet evaporated is therefore
\begin{equation}
	\Omega_{\mathrm{PBH}}
		=
								\frac{M\,n_{\mathrm{PBH}}( t_{0} )}{\rho_{\mathrm{crit}}}
		\approx
								\left( \frac{\beta( M )}{1.03 \times 10^{-8}} \right)
								\left( \frac{h}{0.68} \right)^{\!-2}
								\gamma^{1/2}\,
								\left( \frac{g_{* \irm}}{106.75} \right)^{\!-1/4}\,
								\left( \frac{M}{M_{\odot}} \right)^{\!-1/2}
								\, ,
								\label{eq:omega}
\end{equation}
which is a more precise form of Eq.~\eqref{eq:roughomega}. Since $\beta$ always appears in combination with $\gamma^{1/2}\,g_{* \irm}^{-1/4}\.h^{-2}$\,, we follow Ref.~\cite{Carr:2009jm} in defining a new parameter
\begin{equation}
	\beta'( M )
		\equiv
								\gamma^{1/2}\,
								\left( \frac{g_{* \irm}}{106.75} \right)^{\!-1/4}
								\left( \frac{h}{0.68} \right)^{\!-2}\,\beta( M )
								\, ,
								\label{eq:betaprime}
\end{equation}
where $g_{* \irm}$ and $h$ can be specified very precisely but $\gamma$ is rather uncertain.
\newpage

An immediate constraint on $\beta'( M )$ comes from the limit on the CDM density parameter, $\Omega_{\mathrm{CDM}}\,h^{2} = 0.110 \pm 0.006$ with $h = 0.72$, so the $3 \sigma$ upper limit is $\Omega_{\mathrm{PBH}} < \Omega_{\mathrm{CDM}} < 0.25$ \cite{Dunkley:2008ie}.
This implies
\begin{equation}
	\beta'( M )
		<
								2.04 \times 10^{-18}\,
								\left( \frac{\Omega_{\mathrm{CDM}}}{0.25} \right)
								\left( \frac{M}{10^{15}\,\grm} \right)^{\!1/2}
								\quad
								( M \gtrsim 10^{15}\,\grm )
								\, .
								\label{eq:density}
\end{equation}
However, this relationship must be modified if the Universe ever deviates from the standard radiation-dominated behaviour. The expression for $\beta'( M )$ may also be modified in some mass ranges if there is a second inflationary phase \cite{Green:1997sz} or if there is a period when the gravitational constant varies \cite{Barrow:1996jk} or there are extra dimensions \cite{Sendouda:2006nu}.

Any proposed model of PBH formation must be confronted with constraints in the mass range where the predicted PBH mass function peaks. These constraints are discussed in Sec.~\ref{sec:Constraints} and expressed in terms of the ratio of the current PBH mass density to that of the CDM density:
\begin{equation}
	f
		\equiv
								\frac{\Omega_{\mathrm{PBH}}}{\Omega_{\mathrm{CDM}}}
		\approx
								4.8\,\Omega_{\mathrm{PBH}}
		=
								4.11 \times 10^{8}\,\beta'( M )\,
								\left( \frac{M}{M_{\odot}} \right)^{\!-1/2}
								\label{eq:f}
								\, ,								
\end{equation}
where we assume $\Omega_{\mathrm{CDM}} = 0.21$. We can also write this as
\begin{equation}
	f
		=
								\beta^{\rm eq} / \Omega_{\rm CDM}^{\rm eq}
		\approx
								2.4\.\beta^{\rm eq}
								\, ,
\label{fraction}
\end{equation}
where $\beta^{\rm eq}$ is the PBH mass fraction at matter-radiation equality. This procedure will be applied in Sec.~\ref{sec:RealisticMassFunctions} to two specific models, the axion-like curvaton model and running-mass inflation (specified in detail in the next section). We will also demonstrate the influence of critical collapse, non-sphericity and non-Gaussianity on the PBH dark-matter fraction.

\subsection{Extended Versus Monochromatic Mass Functions}
\label{sec:Extended-versus-monochromatic-mass-functions}

As regards the representation of constraints for extended mass functions, one approach is to integrate the differential mass function $\d n / \d M$ over a mass window of width $M$ at each $M$, giving the continuous function
\begin{equation}
	n( M )
		=
								M \frac{\d n}{\d M}
		=
								\frac{\d n}{\d \ln M}
								\, .
\end{equation}
Here $n( M )$ can be interpreted as the number density of PBHs in the mass range ($M, 2\.M$). One can then define the quantities
\begin{equation}
	\rho( M )
		=
								M^{2} \frac{\d n}{\d M}
								\, ,
								\quad
	f( M )
		=
								\frac{ \rho( M )}{\rho_{\mathrm{CDM}} }
								\, ,
\end{equation}
which correspond to the mass density and dark-matter fraction, respectively, in the same mass range. This is equivalent to breaking the mass up into bins and has the advantage that one can immediately see where most of the mass is. If one knows the expected mass function, one can plot $n( M )_{\rm exp}$ or $\rho( M )_{\rm exp}$ in the same figure as the constraints to see which one is strongest. Alternatively, one can define $n( M )$, $\rho( M )$ and $f( M )$ as integrated values for PBHs with mass larger or smaller than $M$. However, these are only simply related to the functions defined above for a power-law spectrum.

The above representations are problematic if the width of the mass function is less then $M$. Indeed, one might {\it define} an extended mass function as one with a width larger than $M$, in which case we have seen that one can always specify an effective value $f( M )$ at each mass-scale. The situation for monochronatic mass functions is generally more complicated, although it is straightforward if the mass function is a delta function (\ie~exactly monochromatic). The problem arises if it is nearly monochromatic (\ie~with width $\Delta M \ll M$). This is discussed in more detail in Ref.~\cite{Carr:2016hva}.

Although a precisely monochromatic mass spectrum is clearly unphysical, one would only expect the mass function to be \emph{very} extended if the PBHs formed from \emph{exactly} scale-invariant density fluctuations \cite{Carr:1975qj} or from the collapse of cosmic strings \cite{Hawking:1987bn}. In this case, one has
\begin{equation}
	\frac{\d n}{\d M}
		\propto
								M^{-5/2}
								\, ,
								\quad
	n( M )
		\propto
								M^{-3/2}
								\, ,
								\quad
	\rho( M )
		\propto
								f( M )
		\propto
								M^{-1/2}
								\, ,
								\quad
	\beta( M )
		=
								\mathrm{constant}
								\, .
\end{equation}
This is not expected in the inflationary scenario but in most circumstances the spectrum would still be extended enough to have interesting observational consequences, since the constraint on one mass-scale may also imply a constraint on neighbouring scales. We have mentioned that the monochromatic assumption fails badly if PBHs form through critical collapse and the way in which this modifies the form of $\beta( M )$ has been discussed by Yokoyama~\cite{Yokoyama:1998xd}. This will be discussed in more detail in Sec.~\ref{sec:RealisticMassFunctions}.

\section{Specific Models}
\label{sec:SpecificModelsInRelevantRanges}

\noindent
For a large fraction of PBH formation scenarios, an extended feature in the primordial density power spectrum is generic. This leads to a non-monochromatic PBH mass spectrum. As demonstrated in the next section, even an initially peaked spectrum of density perturbation will acquire a significant broadening. In the light of the recent detection of merging black holes in the intermediate-mass range $10\,M_{\odot} < M < 10^{2}\,M_{\odot}$ by the LIGO and Virgo collaboration \cite{Abbott:2016blz, Abbott:2016nmj}, we consider below some models which are capable of producing PBHs in this or one of the other two possible mass intervals. We look first at running-mass inflation and the axion-like curvaton model. These are chosen because their parameters can be tuned so as to give a peak in PBH production in any of the three ranges we would like to investigate. They also have the advantage of not being ruled out by non-Gaussianity effects. We will determine the mass functions explicitly in Sec.~\ref{sec:RealisticMassFunctions} and confront them with recent observational bounds in Sec.~\ref{sec:ExtendedPBH}. We will also briefly review scale-invariant mass functions.

\subsection{Running-Mass Inflation}
\label{sec:Running--Mass-Inflation}

PBH formation in the running-mass model \cite{Stewart:1996ey, Stewart:1997wg} has been intensively studied in Refs.~\cite{Drees:2011yz, Drees:2011hb, Drees:2012sz}; see also Ref.~\cite{Leach:2000ea} for a discussion of constraints and Ref.~\cite{Kuhnel:2015vtw} for an investigation of critical collapse in these models. Perhaps the simplest realisation of this is the inflationary potential
\begin{align}
	\label{pot1}
	V( \phi )
		&=
								V_{0}
								+
								\frac{ 1 }{ 2 }\.m_{\phi}^{2}(\phi)\.\phi^{2}
								\, ,
\end{align}
where $\phi$ is the scalar field and $V_{0}$ is a constant. There exists a plethora of embeddings of this model in various frameworks, such as hybrid inflation \cite{Linde:1993cn}, which lead to different functions $m_{\phi}( \phi )$. These yield distinct expressions for the primordial density power spectra whose variance can be recast into the general form \cite{Drees:2011hb}
\begin{align}
	\big[ \sigma( k ) \big]^{2}
		&\simeq
								\frac{ 8 }{ 81 }\.\Pcal( k_{\star} )
								\bigg(
									\frac{ k }{ k_{\star} }
								\bigg)^{\!\! n( k ) - 1}
								\Gamma\!
								\left(
									\frac{ n_{\srm}( k ) + 3 }{ 2 }
								\right)
								,
								\label{eq:sigma-running-mass}
\end{align}
where the spectral indices $n( k )$ and $n_{\srm}( k )$ are given by
\begin{subequations}
\begin{align}
	n( k )
		&=
								n_{\srm}( k_{\star} )
								-
								\frac{ 1 }{ 2! }\.a\.\ln\!
								\bigg(
									\frac{ k }{ k_{\star} }
								\bigg)
								+
								\frac{ 1 }{ 3! }\.b\.\ln^{2}\!
								\bigg(
									\frac{ k }{ k_{\star} }
								\bigg)
								-
								\frac{ 1 }{ 4! }\.c\.\ln^{3}\!
								\bigg(
									\frac{ k }{ k_{\star} }
								\bigg)
								+
								\ldots
								\, ,
								\label{eq:n-running-mass}
								\displaybreak[1]
								\\[2mm]
	n_{\srm}( k )
		&=
								n_{\srm}( k_{\star} )
								-
								a\.\ln\!
								\bigg(
									\frac{ k }{ k_{\star} }
								\bigg)
								+
								\frac{1}{2}\.b\.\ln^{2}\!
								\bigg(
									\frac{ k }{ k_{\star} }
								\bigg)
								-
								\frac{1}{6}\.c\.\ln^{3}\!
								\bigg(
									\frac{ k }{ k_{\star} }
								\bigg)
								+
								\ldots
								\, ,
\end{align}
\end{subequations}
with real parameters $a$, $b$, and $c$.

As the spectral index and amplitude of the primordial power spectrum at the pivot scale $k_{\star} = 0.002\.{\rm Mpc}^{-1}$ have been measured \cite{Komatsu:2010fb, Planck:2013jfk, Ade:2015lrj} to be $n_{\rm s}( k_{\star} ) \approx 0.96 < 1$ and $\Pcal( k_{\star} ) = \Ocal( 10^{-9} )$, respectively, models without running cannot produce an appreciable PBH abundance. Furthermore, with the measurement of $a = − 0.003 \pm 0.007 \ll 1$ \cite{Ade:2015lrj}, running alone cannot give sufficient increase of the power spectrum at early times. One needs to include at least a running-of-running term, this being subject only to the weak constraint $b \simeq 0.02 \pm 0.02$ \cite{Planck:2013jfk, Ade:2015lrj}. In order to avoid overproduction of PBHs on the smallest scales, a running-of-running-of-running parameter is also needed, so a minimal viable model has all three parameters $a$, $b$ and $c$.

\subsection{Axion-Curvaton Inflation}
\label{sec:Axion--Curvaton-Inflation}

The original curvaton scenario was introduced by Lyth {\it et al.} \cite{Lyth:2001nq, Lyth:2002my}. The model we investigate here is a variant of this and was introduced by Kasuya and Kawasaki \cite{Kasuya:2009up} (\cf~\cite{Kohri:2012yw, Kawasaki:2012wr}). It describes a curvaton moving in an axion or natural inflation-type potential. For a recent study of PBH production in this model, including critical collapse, see Ref.~\cite{Kuhnel:2015vtw}.

In this model, the inflaton $\phi$ is the modulus and the curvaton $\chi$ is related to the phase $\theta$ of a complex superfield $\Phi$. In practice, the inflaton rolls down a potential of the form
\begin{align}
	V( \phi )
		&=
								\frac{ 1 }{ 2 }\.\lambda\,H^{2}\phi^{2}
								\, ,
\end{align}
where $H$ is the Hubble rate and $\lambda$ is a constant derived from combinations of parameters in supergravity theory. Because of its large mass, the inflaton rolls fast towards its minimum $\phi_{\rm min}$. After this, the curvaton becomes well-defined as $\chi = \phi_{\rm min} \theta \sim f \theta$ and this becomes the primary degree of freedom of the superfield. The curvaton is assumed to move in an axion-like potential, similar to that of natural inflation \cite{Freese:1990rb},
\begin{align}
	V_{\chi}
		&=
								\Lambda^{4}
								\left[
									1
									-
									\cos\!
									\left(
										\frac{ \chi }{ f }
									\right)
								\right]
		\simeq
								\frac{1}{2}\.m_{\chi}^{2}\.\chi^{2}
								\, ,
\end{align}
where the last equality holds when $\chi$ is close to its minimum at $0$ and the curvaton mass is $m_{\chi} = \Lambda^{2} / f$. The particular shape of this potential, which preserves the shift-symmetry peculiar to axions, is what makes this curvaton axion-like.

The power spectrum of primordial perturbations is generated by the combined effect of the inflaton and the curvaton,
\vs{-2mm}
\begin{align}
	\Pcal_{\zeta} (k)
		&=
								\Pcal_{\zeta,\mathrm{inf}}( k )
								+
								\Pcal_{\zeta,\mathrm{curv}}( k )
								\, .
\end{align}
The first term is dominant on large scales (small $k$) and the second on small scales (large $k$). The inflaton perturbation is assumed to yield a near scale-invariant spectrum with $\Pcal_{\zeta,\mathrm{inf}}( k ) \simeq 2\times 10^{-9}$, in accordance with CMB observations \cite{Komatsu:2010fb, Planck:2013jfk, Ade:2015lrj}. This contribution should dominate up to at least $k \sim 1\,\mathrm{Mpc}^{-1}$. We define $k_{\crm}$ as the crossing scale at which the curvaton and inflaton contributions to the power spectrum are equal, and $k_{\frm}$ as the scale at which the inflaton reaches its minimum, $\phi_{\rm min} \sim f$, so that the curvaton becomes well-defined. $M_{\crm}$ and $M_{\frm}$ are the horizon masses when these scales cross the horizon. PBHs cannot form before these horizon-crossing times, because the perturbations are too small when $M_{\mathrm{H}}> M_{\crm}$, and no curvaton perturbations exist for $M_{\mathrm{H}} > M_{\frm}$. Here $M_{\frm}$ can be found explicitly from the parameters of the theory and has the value
\begin{align}
	M_{\frm}
		&\approx
								10^{13 - 12 / ( n_{\chi} - 1 )}
								\left(
									\frac{g_{\frm}}{100}
								\right)^{-1/6}
								\left(
									\frac{ k_{\crm} }{\mathrm{Mpc}^{-1} }
								\right)^{-2}
								\left(
									\frac{ \Pcal_{\zeta, \mathrm{curv}}( k_{\frm} ) }{ 2 \times 10^{-3} }
								\right)^{- 2 / ( n_{\chi}- 1 )}
								M_{\odot}
								\, ,
\end{align}
where $n_{\chi}$ is the curvaton spectral index and $g_{\frm}$ is the number of radiative effective degrees of freedom at the scale $k_{\frm}$. Throughout our considerations, we will follow Ref.~\cite{Kawasaki:2012wr} in assuming $k_{\crm} = 1\,\mathrm{Mpc}^{-1}$ and $g_{\frm} = 100$.

PBHs cannot form from the inflationary density perturbations, as these are constrained by CMB observations. By contrast, when the curvaton power spectrum becomes dominant, it can have much more power and still evade the CMB bounds, allowing the production of large PBHs. However, the curvaton perturbations are assumed not to collapse to PBHs before the inflaton has decayed to standard-model particles. Hence PBHs can only form with the minimum mass $M_{\rm min}$, these being produced at or after the curvaton decay time. The exact value for the decay time{\;---\;}and hence the minimum mass $M_{\rm min}${\;---\;}is not known, but it should be smaller than the horizon mass at BBNS ($10^{38}\,\grm$), in order not to interfere with this process, and smaller than $M_{\frm}$ to yield PBH production. In Ref.~\cite{Kawasaki:2012wr}, $M_{\rm min}/M_{\frm} = 10^{-8}$ and $10^{-3} $ are considered, so we will do the same here.

It can be shown that the variance of the density power spectrum due to the curvaton perturbations in a model with an axion-like curvaton is \cite{Kawasaki:2012wr}
\begin{align}
	\sigma_\delta^{2}(M_{H})
		&=
								\frac{8}{81}\,\Pcal_{\zeta,\mathrm{curv}}( k_{\frm} )
								\left[
									\left(
										\frac{ M_{\frm} }{ M_{H} }
									\right)^{(n_{\chi} - 1)/2}
									\gamma\mspace{-1mu}
									\left(
										\frac{ n_{\chi} - 1 }{ 2 }, \frac{ M_{H} }{ M_{\frm} }
									\right)
									+
									E_{1}\!
									\left(
										\frac{ M_{H} }{ M_{H_{0}} }
									\right)
								\right]
								\label{eq:sigma-axion-like}
\end{align}
for a horizon mass $M_{H} > M_{\rm min}$. For $M_{H} < M_{\rm min}$, we assume the curvaton power spectrum which can transform into PBHs is zero. Due to inhomogeneities of the curvaton decay, this is not strictly true. However, as in Ref.~\cite{Kawasaki:2012wr}, we will take this to be a reasonable approximation. The curvaton spectral index is controlled by the parameter $\lambda$:
\begin{align}
	n_{\chi} - 1
		&=
								3
								-
								3\.\sqrt{
										1
										-
										\frac{4}{9} \lambda
									\,} \, .
\end{align}
By setting $\lambda \in (1,\.9 / 4]$, we can obtain a sufficiently blue power spectrum of curvature perturbations for the curvaton to produce PBHs at some scale without violating the CMB constraints. The minimum mass $M_{\rm min}$, defined by the decay time of the curvaton, protects the model from overproducing PBHs at very small scales in spite of the blue power spectrum. The functions $\gamma$ and $E_{1}$ are given by
\begin{subequations}
\begin{align}
	\gamma( a, x )
		&\equiv
								\int_{0}^{x}\d t\;t^{a - 1}\erm^{-t}
								\, ,
								\displaybreak[1]
								\\[2mm]
	E_{1}( x )
		&\equiv
								\int_{x}^{\infty}\d t\;\frac{ \erm^{-t} }{ t }
								\, ,
\end{align}
\end{subequations}
which are the lower incomplete gamma function and the exponential integral, respectively.

\subsection{Scale-invariant Mass Functions}
\label{sec:ScaleInvariant}

\noindent
For a scale-invariant PBH mass function (\ie~with $\beta( M )$ independent of $M$), one has $\Omega_{\rm PBH}( M ) \propto M^{-1/2}$ and so the largest contribution to the dark-matter density comes from the smallest holes. One therefore needs to specify the lower mass cut-off $M_{\rm min}$ and then check that the implied value of $\beta( M )$ on scales above $M_{\rm min}$ does not violate any of the other PBH constraints. For $M_{\rm min} < 10^{15}\,\grm$, the strongest constraint is likely to come from the $\gamma$-ray background limit $\beta( M_{*} ) < 3 \times 10^{-27}$ \cite{Carr:2009jm}.

Cosmic strings produce PBHs with a scale-invariant-mass function, with the lower cut-off being associated with the symmetry-breaking scale \cite{Hogan:1984zb, Hawking:1987bn, Polnarev:1988dh, Garriga:1993gj, Caldwell:1995fu, Cheng:1996du, MacGibbon:1997pu, Hansen:1999su, Nagasawa:2005hv}. Since $\Omega_{\rm PBH}( M )$ decreases with increasing $M$ and $\Omega_{\rm PBH}(10^{15}\,\grm) < 10^{-8}$ from the $\gamma$-ray background limit, such PBHs cannot provide the dark matter unless $M_{\rm min}$ exceeds $10^{15}$g, which seems implausible. On the other hand, if evaporating black holes leave stable Planck-mass relics, these might also contribute to the dark matter. The discussion in Sec.~\ref{sec:ExtendedPBH}.D shows that the $\gamma$-ray background limit excludes relics from providing all of the dark matter unless $M < (15\.\kappa )^{2 / 3} M_{\rm Pl}$.

\section{Extended mass functions: criticality, non-sphericity and non-Gaussianity}
\label{sec:RealisticMassFunctions}

\noindent
The simplest model of PBH formation assumes that the mass spectrum is monochromatic{\;---\;}with mass comparable to the horizon mass at formation{\;---\;}and that the PBHs derive from the collapse of overdensities which are spherical and have a Gaussian distribution. Given the large uncertainties in the PBH formation process and the plethora of models for it, this na{\" i}ve approach has been adopted in many papers. This includes Ref.~\cite{Carr:2009jm}, which discusses the numerous constraints on the PBH abundance as a function of mass on the assumption that the mass spectrum has a width $\Delta M$ of order $M$. As the bounds on the allowed PBH density at each epoch have become more refined, and since PBHs of intermediate mass may even have been observed~\cite{Bird:2016dcv}, a more precise treatment of the formation process is necessary. In this section we therefore go beyond the usual assumptions and attempt a more realistic treatment.

\begin{figure}[t]
	\centering
	\includegraphics[scale=0.89,angle=0]{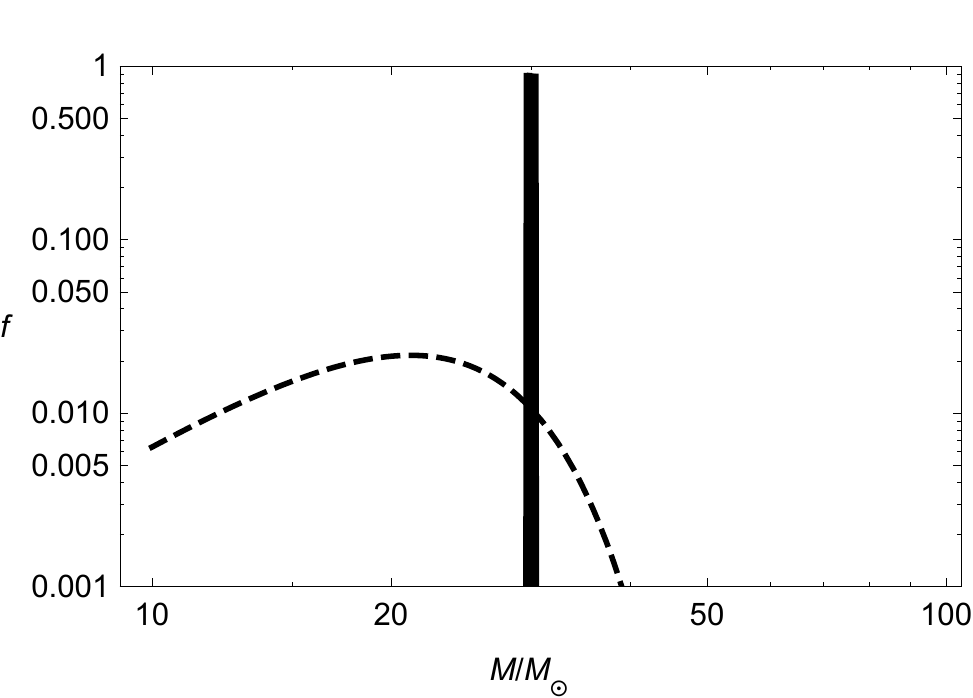}\;
	\includegraphics[scale=0.89,angle=0]{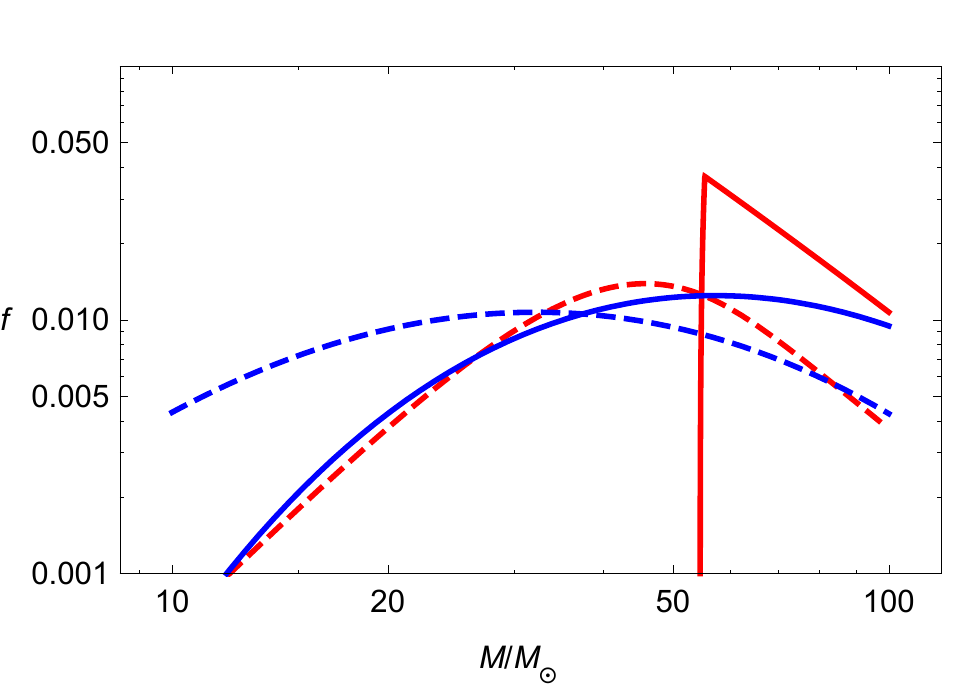}
	\caption{Effect of critical collapse on the fraction $f$ as a function of black-hole mass in units of solar mass 
		for a nearly monochromatic mass function ({\it left panel}), and for axion-like curvaton (red) as well as 
		running-mass inflation (blue) ({\it right panel}). 
		The latter models are specified in Sec.~\ref{sec:SpecificModelsInRelevantRanges}. 
		Solid lines show the PBH abundances for the horizon-mass collapse estimate, 
		dashed lines show the same models with critical collapse included.}
	\label{fig:Critical-collapse}
\end{figure}

\subsection{Monochromaticity}
\label{sec:Monochromaticity}

The monochromatic assumption is a good starting point if the spread in mass is narrow enough, but this is not very likely for most inflationary models which produce PBHs \cite{Clesse:2015wea, Kohri:2012yw, Drees:2011hb}. As reviewed below, although models exist with a narrow spectrum, such as the axion curvaton model \cite{Kawasaki:2012wr} or some phase transition models \cite{Kodama:1982sf, Jedamzik:1999am, Khlopov:2000js}, more realistic treatments{\,---\,}involving critical collapse{\,---\,}yield extended mass functions. This applies even if the PBHs derive from a very narrow feature in the original power spectrum, as illustrated in Fig.~\ref{fig:Critical-collapse}. This can lead to the misinterpretation of observational constraints, since these are mostly derived for monochromatic mass functions. We shall discuss how to treat constraints for extended mass functions in Sec.~\ref{sec:ExtendedPBH} (\cf~a related discussion in Ref.~\cite{Carr:2016hva}).

\subsection{Critical Collapse}
\label{sec:Critical-Collapse}

Early research assumed that a sufficiently large overdensity re-enters the horizon and collapses to a black hole of order the horizon mass $M_{H}$ almost immediately. However, under the assumption of spherical symmetry, it has been shown \cite{Choptuik:1992jv, Koike:1995jm, Niemeyer:1999ak, Gundlach:1999cu, Gundlach:2002sx} that the functional dependence of the PBH mass $M$ on $\delta$ and $M_{H}$ follows the critical scaling relation
\begin{align}
	M
		&=
								k\.M_{H}\.
								\big(
									\delta
									-
									\delta_{\crm}
								\big)^{\gamma}
								\label{eq:M-delta-scaling}
\end{align}
for $\delta > \delta_{\crm}$. The constant $k$, the threshold $\delta_{\crm}$ and the critical exponent $\gamma$ all depend on the nature of the fluid containing the overdensity $\delta$ at horizon-crossing \cite{Musco:2012au}. Careful numerical work \cite{Musco:2004ak, Musco:2008hv, Musco:2012au} has confirmed the scaling law \eqref{eq:M-delta-scaling}. In particular, Fig.~1 of Ref.~\cite{Musco:2008hv} suggests that it applies over more than $10$ orders of magnitude in density contrast.

Soon after the first studies of critical collapse \cite{Choptuik:1992jv, Koike:1995jm}, its application to PBH formation was studied and incorporated in concrete models \cite{Niemeyer:1998ac, Green:1999xm}. The conclusion was that the horizon-mass approximation was still reasonably good. However, this conclusion depended on the assumption that the mass function would otherwise be monochromatic. As shown in \cite{Kuhnel:2015vtw} for a variety of inflationary models, when a realistic model of the power spectrum underlying PBH production is used, the inclusion of critical collapse can lead to a significant shift, lowering and broadening of the PBH mass spectra{\;---\;}sometimes by several orders of magnitude.

Regarding Eq.~\eqref{eq:M-delta-scaling}, it has been shown that the critical exponent $\gamma$ is independent of the perturbation profile \cite{Neilsen:1998qc, Musco:2012au}, though $\delta_{\rm c}$ and $k$ may depend on this. Throughout this work we shall follow the pioneering work of Ref.~\cite{Carr:1975qj} and later Ref.~\cite{Green:2004wb} in applying the Press--Schechter formalism \cite{1974ApJ...187..425P} for spherical collapse. As a first approximation, we assume a Gaussian perturbation profile,
\begin{align}
	\Pcal( \delta )
		&\equiv
								\frac{ 1 }{ \sqrt{2 \pi \sigma^{2}\,} }\,
								\exp\!
								\left(
									- \frac{ \delta^{2} }{ 2\.\sigma^{2} }
								\right)
								,
								\label{eq:P-Gaussian}
\end{align}
which accords well with current CMB measurements \cite{Ade:2015ava}. As detailed below, this should be generalized to include simple non-Gaussian profiles, as even the slight non-Gaussianities permitted by observation may alter the final PBH abundance \cite{Young:2013oia, Young:2015kda}. The quantity $\sigma$ is the variance of the primordial power spectrum of density perturbations generated by the model of inflation. In radiation-dominated models, which are the focus of this paper, repeated studies have shown that $\gamma \simeq 0.36$ \cite{Koike:1995jm, Niemeyer:1999ak, Musco:2004ak, Musco:2008hv, Musco:2012au} and $\delta_{\rm c} \simeq 0.45$ \cite{Musco:2004ak, Musco:2008hv, Musco:2012au}. In accordance with Ref.~\cite{Niemeyer:1997mt}, we set $k = 3.3$.

As mentioned in Sec.~\ref{sec:ModelsInRelevantRanges}, a convenient measure of how many PBHs are produced is the ratio of the PBH energy density to the total energy density at PBH formation. Using the Press--Schechter formalism, we can express this as
\begin{align}
	\beta
		&=
								\int_{\delta_{\crm}}^{\infty}
								\d \delta\;
								k
								\big(
									\delta
									-
									\delta_{\crm}
								\big)^{\!\gamma}_{}\,
								\Pcal( \delta )
		\approx
								k\.\sigma^{2\gamma}\,
								{\rm erfc}
								\bigg(
									\frac{ \delta_{\crm} }{ \sqrt{2\,}\.\sigma }
								\bigg)
								\, ,
								\label{eq:Beta_normalisation}
\end{align}
where we assume $\sigma \ll \delta_{\crm}$. We have numerically confirmed the validity of this approximation for our purposes but some subtleties are involved here. These concern the validity of the Press--Schechter formalism, the use of the density rather than curvature power spectrum, and the upper integration limit. We discuss these points more thoroughly below but none of them changes the main signatures of critical collapse, which are the broadening, lowering and shifting. Nevertheless, these small effects should be accounted for in obtaining precise constraints on inflationary models.

Following Ref.~\cite{Niemeyer:1997mt}, we next derive the PBH initial mass function $g$. We define this as the black-hole number $\d N_{\rm PBH}$ per normalised mass interval $\d \mu$, where $\mu \equiv M / ( k\.M_{H} )$, within each collapsing horizon:\footnote{Note the slight difference in the definition of the initial mass function compared to the one in Ref.~\cite{Niemeyer:1997mt}. The latter is expressed in terms of a logarithmic derivative, whereas Eq.~\eqref{eq:PBH-IMF-definition} involves an ordinary derivative, corresponding to an extra inverse power of $\mu$ on the \rhs}
\begin{align}
	g
		&\equiv
								\frac{ \d N_{\rm PBH} }{ \d \mu }
		\equiv
								\frac{ 1 }{ \beta }\;
								\Pcal( \delta[ \mu ] )\.
								\frac{ \d \delta[ \mu ]}{ \d \mu }
		\simeq
								\frac{\mu^{\frac{ 1 }{ \gamma } - 1}\,
 									\exp{\!
										\left[
											-
											\left(
												\delta
												+
												\mu^{\frac{ 1 }{ \gamma }}
												\right)^{2} /
											\,\big( 2 \sigma ^{2} \big)
										\right]
										}
									}
									{ \sqrt{2 \pi\,}\.\gamma\.\sigma\,
										{\rm erfc}\!
										\left(
											\frac{\delta}
											{\sqrt{2\,} \sigma }
										\right)
								}
								\, .
								\label{eq:PBH-IMF-definition}
\end{align}
Here the factor $1 / \beta$ is required to normalise $g$ so that $\int\d\mu\;g( \mu ) = 1$. In deriving Eq.~\eqref{eq:PBH-IMF-definition}, we have used the Gaussian profile \eqref{eq:P-Gaussian} for the amplitude of the fluctuations, the last relation holding for $\sigma \ll \delta_{\crm}$. When we apply the critical-collapse scenario in the subsequent sections, we will follow the procedure outlined at the end of Sec.~II in Ref.~\cite{Kuhnel:2015vtw}. In particular, given a certain horizon mass, this means that we must consider how the PBH mass distribution is spread around the initial mass function $g$. Each contribution has to be evolved from the time of PBH formation to the time of radiation-matter equality, so that the dark-matter fraction $f$, given by Eq.~\eqref{fraction}, can be evaluated by summing over the individual contributions from each horizon mass.

Examples of how critical collapse affects the abundance and mass distribution of PBHs can be seen in Fig.~\ref{fig:Critical-collapse}. The left panel shows the application of the above scheme for critical collapse to a nearly monochromatic feature in the initial density perturbations. As can be seen, the resulting mass function is far from monochromatic and yields PBHs over a wide range of masses. In the right panel of Fig.~\ref{fig:Critical-collapse} our scheme has been applied to initial perturbation spectra from two inflationary models, the axion-like curvaton model and a running-mass model, the details of which have been discussed in Sec.~\ref{sec:SpecificModelsInRelevantRanges}. In all cases, the change in shape and mass range due to critical collapse is clearly visible. This is particularly true for the axion-like curvaton model or the critical-collapse version of the monochromatic function where the slope for masses smaller then its initial peak, is entirely due to critical collapse. Fig.~7 of Ref.~\cite{Carr:2016hva} shows the same behaviour, \ie~the power-law tails towards lower masses are equivalent.

\subsection{Non-sphericity}
\label{sec:Non-sphericity}

The above results rely on the assumption of spherical collapse. The inclusion of non-sphericity is significantly more complicated and has not been subject to extensive numerical studies of the kind in Ref.~\cite{Musco:2012au}. Inspired by related work on gravitational collapse in the context of galactic halo formation. where it has been known for a long time (\cf~\cite{Sheth:1999su}) that non-zero ellipticity leads to possibly large effects, Ref.~\cite{Kuhnel:2016exn} shows that this also holds for PBH mass spectra. One essential consequence is that the threshold value is increased and can generically be approximated as
\begin{align}
	\frac{ \delta_{\rm ec} }{ \delta_{\crm} }
		&\simeq
								1
								+
								\kappa\mspace{-1mu}
								\left(
									\frac{ \sigma^{2} }{ \delta_{\crm}^{2} }
								\right)^{\!\!\gamma}
								,
								\label{eq:Sheth-Mo-Tormen}
\end{align}
with $\delta_{\crm}$ being the threshold value for spherical collapse and $\sigma^{2}$ the amplitude of the density power spectrum at the given scale. In Ref.~\cite{Sheth:1999su} the above result was derived and numerically confirmed for a limited class of cosmologies, mostly relevant to structure formation, where $\kappa$ and $\gamma$ were found to be $0.47$ and $0.62$, respectively. In particular, this does not include the case of ellipsoidal collapse in a radiation-dominated model, which is most relevant for PBH formation.

In Ref.~\cite{Kuhnel:2016exn} it was argued that a relation of the form of Eq.~\eqref{eq:Sheth-Mo-Tormen} should hold for ellipsoidal gravitational collapses in arbitrary environments. Schematically, the argument goes as follows: The collapse starts along the smallest axis and thereafter the longer axes collapses faster than linearly \cite{Bond:1993we}. The mass dependence of the overdensity $\delta( M )$ suggests that the density perturbation in the primary collapsing sphere{\,---\,}with radius equal to the shortest axis{\,---\,}will be smaller by $\delta( \Delta M )$. Here $\Delta M$ accounts for the difference in mass $M$ of a sphere and an ellipsoid. By considering Gaussian-distributed overdensities, it can be shown that the expectation values for the shape of overdensities are \cite{Doroshkevich1970, Bardeen:1985tr, Bond:1993we}
\vs{-0.3mm}
\begin{align}
	\langle e \rangle
		&=
								\frac{3\.\sigma}{\sqrt{10\pi\,}\.\delta}
								\, ,
	\quad\quad\quad\quad
	\langle p \rangle
		=
								0
								\, ,
								\label{eq:<e><p>}
\end{align}
where $e$ is the ellipticity and $p$ the prolateness, which runs from $p = e$ in the maximally prolate case to $p = -\.e$ in the maximally oblate case. Since the collapse is initiated along the shortest axis, it may be compared to that of the largest sphere contained within it. The volume of the ellipsoid is then $V_{\!e} = V_{\!s}\;( 1 + 3 e ) / \sqrt{1 - 3 e\;}$. Taking the ellipsoid to be of uniform density, combined with the demand that the density threshold should be exceeded in the enclosed sphere, leads to an increase in mass $M_{e} = M_{s}\.V_{\!e} / V_{\!s}$. As the density contrast associated with a given mass roughly scales as $\delta( M ) \sim M^{2 / 3}$ in the PBH case \cite{Carr:1975qj}, to first order in the ellipticity this leads to Eq.~\eqref{eq:Sheth-Mo-Tormen} with $\kappa = 9 / \sqrt{10\.\pi\,}$ and $\gamma = 1 / 2$.

In more realistic situations, these values will not be exact and a thorough numerical investigation is needed to precisely determine the change of the threshold for fully relativistic non-spherical collapse. In particular, the above derivation assumes a uniform density in the ellipsoid, whereas a density profile with higher density in the central regions seems more realistic. This should lead to a less pronounced effect. However, the effect will always be an increase in the threshold, leading to a general suppression of the mass spectrum.

The left panel of Fig.~\ref{fig:Non-Sphericity-and-Non-Gaussianity} shows the effect of non-sphericity on the PBH mass fraction $f$, which is given as a function of black-hole mass for the axion-like curvaton model (red) and the running-mass model (blue). (These will be discussed in detail in Sec.~\ref{sec:ExtendedPBH}.) For both models, the solid lines involve only critical collapse, while the dotted lines also include the effect of ellipticity according to our na{\"i}ve model with $\kappa = 9 / \sqrt{10\.\pi\,}$ and $\gamma = 1 / 2$. One can see a significant global shift downwards. This general behaviour is expected because non-spherical effects raise the formation threshold, making it harder for PBHs to from.

Although the exact amount of suppression due to non-sphericity is not known, the functional form of the mass spectrum is essentially unchanged. Also there is degeneracy with the effects of other parameters. Therefore we will not include this effect explicitly when comparing models with observational constraints in Sec.~\ref{sec:ExtendedPBH}. Nevertheless, if one wants to use PBH constraints on concrete inflationary theories, the suppression due to non-sphericity should be properly accounted for. In fact, for precise constraints, numerical relativistic modelling of ellipticity is required.

Note also that if the overdensities are non-Gaussian, the ellipticity is no longer given by Eq.~\eqref{eq:<e><p>} and one should pay attention to the interplay between these two effects. We will not take this into account here, as an exact knowledge of the non-Gaussianities is needed, but the main effect will again be to change the amount by which the amplitude is shifted. These uncertainties will hence be degenerate with uncertainties in the ellipticity effects.

\subsection{Non-Gaussianity}
\label{sec:Non-Gaussianity}

As PBHs form from the extreme high-density tail of the spectrum of fluctuations, their abundance is acutely sensitive to non-Gaussianities in the density-perturbation profile \cite{Young:2013oia, Bugaev:2013vba}. For certain models{\;---\;}such as the hybrid waterfall or simple curvaton models \cite{Bugaev:2011wy, Bugaev:2011qt, Sasaki:2006kq}{\;---\;}it has even been shown that no truncation of non-Gaussian parameters can be made to the model without changing the estimated PBH abundance \cite{Young:2013oia}. However, non-Gaussianity induced PBH production can have serious consequences for the viability of PBH dark matter. PBHs produced with non-Gaussianity lead to isocurvature modes that could be detected in the CMB \cite{Young:2015kda,Tada:2015noa}. With the current Planck exclusion limits \cite{Ade:2015lrj}, this leads to a constraint on the non-Gaussianity parameters for a PBH-producing theory of roughly $| f_{\rm NL} |, | g_{\rm NL} | < 10^{-3}$. For theories like the curvaton and hybrid inflation models \cite{Linde:1993cn, Clesse:2015wea}, this leads to the immediate exclusion of PBH dark matter, as the isocurvature effects would be too large. Ref.~\cite{Chisholm:2005vm} claims this isocurvature production is generic to PBHs, since they represent very large perturbations. However, the analysis was probably not appropriate for PBHs from a generic source. It is important to note that these constraints can change somewhat if the non-Gaussianities are non-local or non-scale-invariant. They are also weakly dependent on the PBH mass, so care should be taken in making definite statements about particular theories when the magnitude of the non-Gaussianities lies close to the bound (see Ref.~\cite{Young:2015kda} for details).

Even if PBHs are produced in the multi-field models, they do not give isocurvature modes on the CMB scale if the CMB and PBH scale are sufficiently decoupled (so that one effectively has a single-field model on the CMB scale). This is because the CMB-scale isocurvature modes are caused by the non-Gaussian correlation between the CMB and PBH scales. Although the non-Gaussian constraints $| f_{\rm NL} |$, $| g_{\rm NL} | < 10^{-3}$ apply in the isocurvature case, these parameters should be evaluated as a correlation between the CMB and PBH scales and this is generally unrelated to the values of $f_{\rm NL}$ and $g_{\rm NL}$ on the CMB scale \cite{Tada:2015noa}.

In order to be realistic, non-Gaussianities should be taken properly into account when considering a model for PBH production. If a certain model with a manifest inflationary origin is considered, the non-Gaussianity parameters should first be obtained. If their values are higher than the above bound, the model is already excluded as a producer of PBH dark matter. In fact, for a realistic treatment of PBH dark-matter production from an inflationary model, this should be the first constraint to consider, as no further investigation of the model is necessary if the non-Gaussianity is too large. If it falls below this limit, it should still be taken into account when calculating abundances. Examples of how this is done in practice can be found in Refs.~\cite{Young:2013oia, Young:2015kda}.

\begin{figure}
	\centering
	\includegraphics[scale=0.89,angle=0]{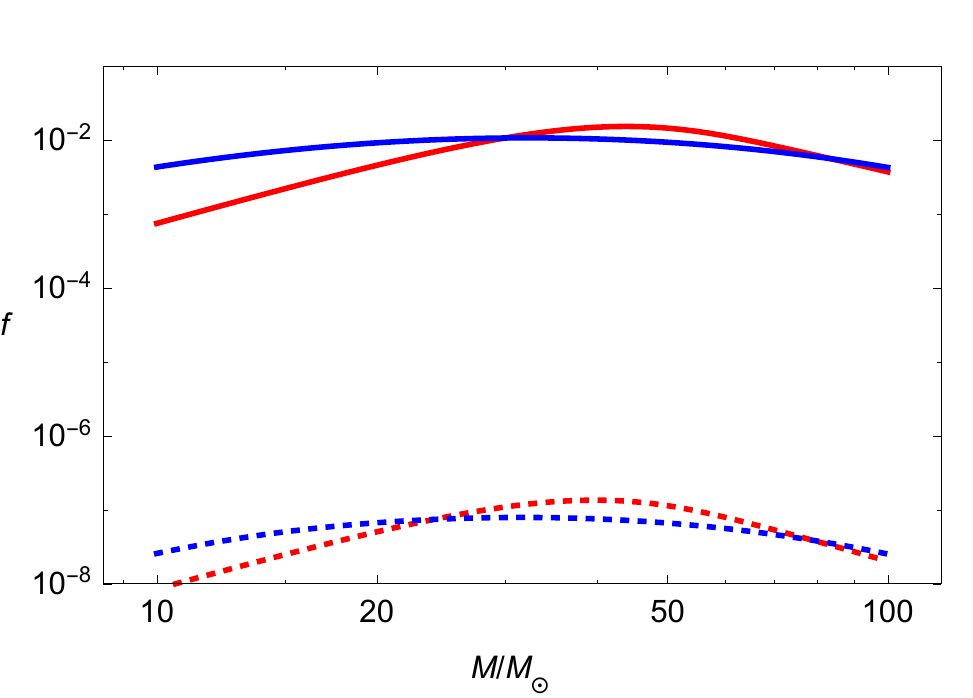}\;
	\includegraphics[scale=0.89,angle=0]{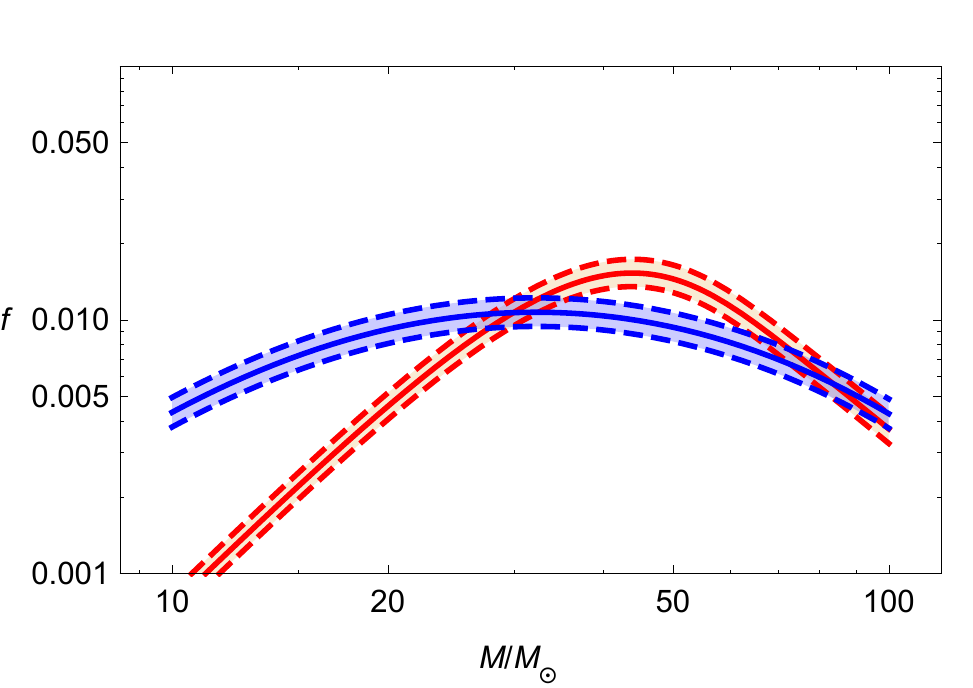}
	\caption{Fraction $f$ as a function of black-hole mass in units of solar mass for axion-like curvaton (red) 
		as well as running-mass inflation (blue); the models are specified in Sec.~\ref{sec:ExtendedPBH}. 
		{\it Left panel}: Effect of non-sphericity 
		(dotted lines), with parameters $\kappa = 9 / \sqrt{10\.\pi\,}$ and 
		$\gamma = 1 / 2$ (see Eq.~\eqref{eq:Sheth-Mo-Tormen}).
		{\it Right panel}: Effect of non-Gaussianity 
		(dot-dashed curves), where we chose 
		$f_{\rm NL} = +\,0.005$ (lower curve) and $f_{\rm NL} = -\,0.005$ (upper curve). 
		In both cases the parameter choices are made for illustrative purpose (see main text for details). 
		Critical collapse 
		is assumed throughout the plots.}
	\label{fig:Non-Sphericity-and-Non-Gaussianity}
\end{figure}

We show an example of the effect of non-Gaussianity in the right panel of Fig.~\ref{fig:Non-Sphericity-and-Non-Gaussianity} for the axion-like curvaton (red) and running-mass inflation (blue) models (to be specified in Sec.~\ref{sec:ExtendedPBH}). Again, the solid curves include only critical collapse, while the dot-dashed curves are for $f_{\rm NL} = \pm\,0.005$. Here, the lower and upper curves correspond to the plus and minus signs, respectively. We have checked that the inclusion of $g_{\rm NL}$ does not have a large effect. The chosen values are of course not precisely accurate for these models; rather they demonstrate the qualitative effect of non-Gaussianity five times larger than the allowed values. In general, the effect is similar to that seen for ellipticity in the left panel of Fig.~\ref{fig:Non-Sphericity-and-Non-Gaussianity}. However, for reasonable values of the non-Gaussianity, the effect is much smaller.

To obtain more precise results, the full nature of the non-Gaussianity should be accounted for. In this work, however, rather than focussing on particular models, we will consider the possibility of non-constrained windows for PBHs to comprise all of the dark matter. We will therefore neglect non-Gaussian effects in our subsequent analysis. More importantly, although not visible directly in our plots, constraints from non-Gaussianity-induced isocurvature must also be considered. This excludes at the outset the production of PBH dark matter in multi-field models. However, the two models discussed in this paper are not affected by this claim: our running-mass model is not multi-field and our axion-curvaton model does not produce curvaton fluctuations on the CMB scale.

\subsection{Miscellaneous Caveats}
\label{sec:Miscellaneous-Caveats}

In addition to the issues mentioned above, some more technical issues concerning the PBH mass spectrum expected from inflationary models have been discussed in the literature. However, none of them are expected to lead to effects which are quantitatively large.

First, we have chosen to use the Press--Schechter formalism for obtaining the mass spectrum from the perturbations. Alternatively, one could calculate the mass fraction using peaks theory \cite{Bardeen:1985tr}. Recently, there has been some discussion \cite{Young:2014ana} of whether these two formalisms predict different values for $f$. If precise constraints on inflationary models are to be obtained from PBH production, this issue should be resolved. However, the signature of this effect would not be a shift or broadening, so the critical collapse effects would be distinguishable from this. In addition, the difference will presumably be much less than the uncertainty in the non-spherical collapse situation. As the issue is currently unresolved, we use Press--Schechter here but attention should be paid to this in the future.

A second (related) subtlety concerns the cloud-in-cloud problem \cite{Jedamzik:1994nr}, which involes the overcounting of small PBHs contained in larger PBHs. This would lead to suppression at the low-mass end of the spectrum. This might counteract the effect of critical collapse but would not occur for spectra deriving from very localised features in the perturbation spectrum. How to account for this is not settled and it is better addressed using peaks theory. Here we will ignore this issue but for precise constraints it should be dealt with properly.

Third, there is the claim \cite{Carr:1974nx} that an overdense region represents a separate closed Universe rather than a part of our Universe if $\delta$ exceeds $1$. In integral \eqref{eq:Beta_normalisation} we have extended the upper integration above $\delta = 1$, in contrast to what was done in Ref.~\cite{Niemeyer:1997mt}. However, Ref.~\cite{Kopp:2010sh} claims that there is no separate-Universe constraint. This is because the meaning of the density perturbation needs to be specified very carefully on large scales: $\delta$ necessarily goes to zero on the separate-univese scale, even though the curvature perturbation diverges. A subsequent discussion \cite{Carr:2014pga} agrees with this conclusion but stresses that the separate-Universe scale is still interesting because it relates to the maximum mass of a PBH forming at any epoch. In any case, the integrand for large values of $\delta$ is so small that this does not make much difference in practice.

Fourth, there are in principle two choices of power spectra from the inflationary models: the curvature power spectrum $\zeta$ and the density power spectrum $\delta$. We choose the latter as this seems to be more accurate for an in-depth discussion (\cf~\cite{Young:2014ana}). This might also link with the separate-Universe and cloud-in-cloud issues \cite{Kopp:2010sh, Harada:2013epa}. Furthermore, Young {\it et al.} \cite{Young:2014ana} have argued that using $\zeta$ instead of $\delta$ to calculate the PBH abundance may yield $O(1)$ errors due to the spurious influence of super-horizon modes.

Fifth, when we evolve our PBH densities through the radiation-dominated epoch, we use a simplified model of cosmic expansion, assuming complete radiation-domination until matter-radiation equality. A more refined treatment should be applied if we are trying to exclude an inflationary model on account of the overproduction of dark matter. However, for the purposes of this paper, this assumption has very little impact. Also this type of modelling would be problematic if one produced more PBHs than there is cold dark matter, as this would change the time of matter-radiation equality, leading to other problems. Careful consideration of this effect may be needed when considering otherwise unconstrained models for the production of evaporating PBHs.

Sixth, once produced, PBHs not only lose mass through Hawking radiation but can also grow by accreting matter and/or radiation or by merging with other PBHs. While Hawking radiation is completely negligible for intermediate-mass PBHs, their growth can be very important in the matter-dominated epoch \cite{Carr:1974nx,1978ApJ...219.1043B,1978ApJ...225..237B}. For instance, it has been conjectured that PBHs with mass of $10^{2} -10^{4}\,M_{\odot}$ could provide seeds for the supermassive black holes of up to $10^{10}\,M_{\odot}$ in the centers of galaxies \cite{2012NatCo...3E1304G}. However, this involves a growth of many orders of magnitude and careful numerical integration is required to study this, allowing for the dilution of the PBHs due to cosmic expansion and the merger of the smaller ones originating from critical collapse. The clustering of PBHs will also have significant effects on their merger rates \cite{Meszaros:1975ef, Carr:1975qj, Meszaros:1980bf}. In particular, Chisholm~\cite{Chisholm:2005vm} showed that the clustering would produce an inherent isocurvature perturbation and used this to constrain the viability of PBHs as dark matter. Later he studied the effect of clustering on mergers \cite{Chisholm:2011kn} and found that these could dominate over evaporation, causing PBHs with mass below $10^{15}\,\grm$ to combine and form heavier long-lived black holes rather than evaporating. So far, no compelling study of this effect has been carried out for a realistic mass spectrum, so we will not include it in our discussion below.

\section{Summary of constraints on monochromatic non-evaporated black holes}
\label{sec:Constraints}

\begin{figure}
	\begin{center}
	\includegraphics{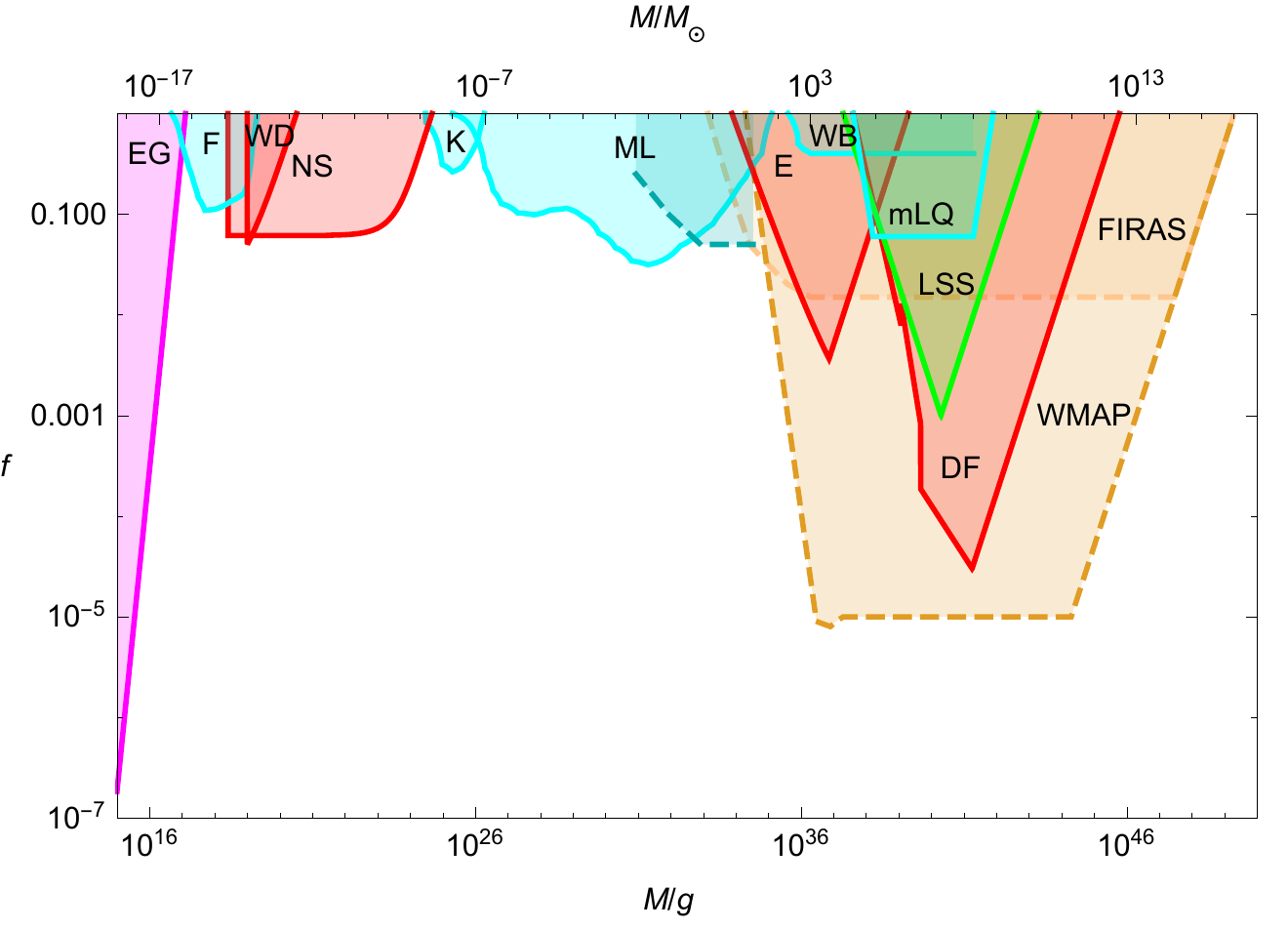}
	\end{center}
	\caption{
		Constraints on $f( M )$ for a variety of evaporation (magenta), dynamical (red), lensing (cyan), 
		large-scale structure (green) and accretion (orange) effects associated with PBHs. 
		The effects are extragalactic $\gamma$-rays from evaporation (EG) \cite{Carr:2009jm},
		femtolensing of $\gamma$-ray bursts (F) \cite{Barnacka:2012bm}, 
		white-dwarf explosions (WD) \cite{2015PhRvD..92f3007G},
		neutron-star capture (NS) \cite{Capela:2013yf}, 
		Kepler microlensing of stars (K) \cite{Griest:2013aaa}, 
		MACHO/EROS/OGLE microlensing of stars (ML) \cite{Tisserand:2006zx, Novati:2013fxa}
		and quasar microlensing (broken line) (ML) \cite{2009ApJ...706.1451M}, 
		survival of a star cluster in Eridanus II (E) \cite{Brandt:2016aco}, 
		wide-binary disruption (WB) \cite{Quinn:2009zg}, 
		dynamical friction on halo objects (DF) \cite{Carr:1997cn}, 
		millilensing of quasars (mLQ) \cite{Wilkinson:2001vv}, 
		generation of large-scale structure through Poisson fluctuations (LSS) \cite{Afshordi:2003zb}, 
		and accretion effects (WMAP, FIRAS) \cite{Ricotti:2007au}. 
		Only the strongest constraint is usually included in each mass range, 
		but the accretion limits are shown with broken lines since they are are highly model-dependent. 
		Where a constraint depends on some extra parameter which is not well-known, 
		we use a typical value. Most constraints cut off at high $M$ due to the incredulity limit. 
		See the original references for more accurate forms of these constraints.}
	\label{fig:large}
\end{figure}

\noindent
We now review the various constraints associated with PBHs which are too large to have evaporated yet, updating the equivalent discussion which appeared in Carr \textit{et al.}~\cite{Carr:2009jm}. All the limits assume that PBHs cluster in the Galactic halo in the same way as other forms of CDM. In this case, the fraction $f( M )$ of the halo in PBHs is related to $\beta'( M )$ by Eq.~\eqref{eq:f}. Our limits on $f( M )$ are summarised in Fig.~\ref{fig:large}, which is an updated version of Fig.~8 of Ref.~\cite{Carr:2009jm}. A list of approximate formulae for these limits is given in Tab.~\ref{tab:ConstraintSummary}. Both Fig.~\ref{fig:large} and Tab.~\ref{tab:ConstraintSummary} are intended merely as an overview and are not exact. A more precise discussion can be found in the original references. Many of the constraints depend on other physical parameters, not shown explicitly. In general, we show only the most stringent constraints in each mass range, although constraints are sometimes omitted when they are contentious. Further details of these limits and similar figures can be found in other papers: for example, Tab.~1 of Josan \textit{et al.} \cite{Josan:2009qn}, Fig.~4 of Mack \textit{et al.} \cite{Mack:2006gz}, Fig.~9 of Ricotti \textit{et al.} \cite{Ricotti:2007au}, Fig.~1 of Capela \textit{et al.} \cite{Capela:2013yf} and Fig.~1 of Clesse \& Garcia-Bellido \cite{Clesse:2016vqa}. We group the limits by type and discuss those within each type in order of increasing mass. Since we are also interested in the mass ranges for which the dark-matter fraction is small, where possible we express each limit in terms of an analytic function $f_{\mathrm{max}}( M )$ over some mass range. We do not treat Planck-mass relics, since the only constraint on these is that they must have less than the CDM density, but we do discuss them further in Sec.~\ref{sec:ExtendedPBH}.

\subsection{Evaporation Constraints}

A PBH of initial mass $M$ will evaporate through the emission of Hawking radiation on a timescale $\tau \propto M^{3}$ which is less than the present age of the Universe for $M$ less than $M_{*} \approx 5 \times 10^{14}$\,g \cite{Carr:2016hva}. PBHs with $M > M_{*}$ could still be relevant to the dark-matter problem, although there is a strong constraint on $f( M_{*} )$ from observations of the extragalactic $\gamma$-ray background \cite{Page:1976wx}. Those in the narrow band $M_{*} < M < 1.005\,M_{*}$ have not yet completed their evaporation but their current mass is below the mass $M_{q} \approx 0.4\,M_{*}$ at which quark and gluon jets are emitted. For $ M > M_{\mathrm c}$, there is no jet emission.

For $M > 2 M_{*}$, one can neglect the change of mass altogether and the time-integrated spectrum $\drm N^{\gamma}/ \drm E$ of photons from each PBH is just obtained by multiplying the instantaneous spectrum $\drm \dot{N}^{\gamma}/ \drm E$ by the age of the Universe $t_{0}$. From Ref.~\cite{Carr:2009jm} this gives
\begin{equation}
	\frac{\drm N^\gamma }{\drm E}
		\propto
				\begin{cases}
						E^{3}\,M^{3}
					& ( E < M^{-1} )\, , \\
						E^{2}\,M^{2}\,\erm^{-E M}
					& ( E > M^{-1} )
					\, .
				\end{cases}
\end{equation}
This peaks at $E \sim M^{-1}$ with a value independent of $M$. The number of background photons per unit energy per unit volume from all the PBHs is obtained by integrating over the mass function:
\begin{equation}
	\Ecal( E )
		=
								\int_{M_{\mathrm{min}}}^{M_{\mathrm{max}}} \!\drm M\,
								\frac{\drm n}{\drm M}\,
								\frac{\drm N^\gamma }{\drm E}(m,E)
								\, ,
\end{equation}
where $M_{\mathrm{min}}$ and $M_{\mathrm{max}}$ specify the mass limits. For a monochromatic mass function, this gives
\begin{equation}
	\Ecal( E )
		\propto
			f( M ) \times
				\begin{cases}
						E^{3}\,M^{2}
					& ( E < M^{-1} )\, ,\\
						E^{2}\,M\,\erm^{-E M}
					& ( E > M^{-1} )\, ,
				\end{cases}
\end{equation}
and the associated intensity is
\begin{equation}
	I( E )
		\equiv
			\frac{ c \, E \, \Ecal( E )}{4 \pi} \propto f( M ) \times
				\begin{cases}
						E^{4}\,M^{2}
					& ( E < M^{-1} )\, ,\\
						E^{3}\,M\,\erm^{- E M}
					& ( E > M^{-1} )\, ,
				\end{cases}
\end{equation}
with units $ \mathrm s^{-1}\, \mathrm{sr}^{-1}\,\mathrm{cm}^{-2}$. This peaks at $E \sim M^{-1} $ with a value $I^{\mathrm{max}} ( M ) \propto f( M ) M^{-2}$. The observed extragalactic intensity is $I^\mathrm{obs} \propto E^{-(1+\epsilon)}\propto M^{1+\epsilon} $ where $\epsilon$ lies between $0.1$ (the value favoured in Ref.~\cite{Sreekumar:1997un}) and $ 0.4 $ (the value favoured in Ref.~\cite{Strong:2004ry}). Hence putting $I^{\mathrm{max}}( M ) \le I^\mathrm{obs} ( M ) $ gives \cite{Carr:2009jm}
\begin{equation}
	f( M )
		\lesssim
				2 \times 10^{-8}\,
				\left(
					\frac{ M }{ M_{*} }
				\right)^{\!\!3 + \epsilon}
		\quad
				( M > M_{*} = 5 \times 10^{14} \grm )
				\; .
				\label{photon2}
\end{equation}
In Fig.~\ref{fig:large} we plot this constraint for $\epsilon = 0.2$. The Galactic $\gamma$-ray background constraint could give a stronger limit \cite{Carr:2016hva} but this requires the mass function to be extended and depends sensitively on its form, so we do not discuss it here. The reionising effects of $10^{16}$ -- $10^{17}\,$g PBHs might also be associated with interesting constraints \cite{Belotsky:2014twa}.

\subsection{Lensing Constraints}
\label{sec:Lensing-Constraints}

Constraints on MACHOs with very low $M$ come from the femtolensing of $\gamma$-ray bursts. Assuming the bursts are at a redshift $z \sim 1$, early studies~\cite{Marani:1998sh,Nemiroff:2001bp} excluded $f = 1$ in the mass range $10^{-16}$ -- $10^{-13}\,M_{\odot}$ but more recent work~\cite{Barnacka:2012bm} gives a limit which can be approximated as
\begin{equation}
	f( M )
		<
				0.1 \quad ( 5 \times 10^{16} \grm < M < 10^{19}\,\grm )
				\, .
				\label{femto}
\end{equation}
The precise form of this limit is shown is Fig.~\ref{fig:large}.

Microlensing observations of stars in the Large and Small Magellanic Clouds probe the fraction of the Galactic halo in MACHOs of a certain mass range \cite{Paczynski:1985jf}. The optical depth of the halo towards LMC and SMC, defined as the probability that any given star is amplified by at least $1.34$ at a given time, is related to the fraction $f$ by
\begin{equation}
	\tau^{(\mathrm{SMC})}_{\mathrm L}
		=
				1.4\,\tau^{(\mathrm{LMC})}_{\mathrm L}
		=
				6.6 \times 10^{-7}\,f
\end{equation}
for the S halo model \cite{Alcock:2000ph}. Although the initial motivation for microlensing surveys was to search for brown dwarfs with $0.02\,M_{\odot} < M < 0.08\,M_{\odot}$, the possibility that the halo is dominated by these objects was soon ruled out by the MACHO experiment \cite{Alcock:2000ke}. However, MACHO observed $17$ events and claimed that these were consistent with compact objects of $M \sim 0.5\,M_{\odot}$ contributing $20\,\%$ of the halo mass \cite{Alcock:2000ph}. This raised the possibility that some of the halo dark matter could be PBHs formed at the QCD phase transition \cite{Jedamzik:1996mr,Widerin:1998my, Jedamzik:1999am}. However, later studies suggested that the halo contribution of $M \sim 0.5\,M_{\odot}$ PBHs could be at most 10\%~\cite{Hamadache:2006fw}. The EROS experiment obtained more stringent constraints by arguing that some of the MACHO events were due to self-lensing or halo clumpiness \cite{Tisserand:2006zx} and excluded $ 6 \times 10^{-8}\,M_{\odot} < M < 15\,M_{\odot}$ MACHOs from dominating the halo. Combining the earlier MACHO \cite{Allsman:2000kg} results with the EROS-I and EROS-II results extended the upper bound to $ 30\,M_{\odot}$ \cite{Tisserand:2006zx}. The constraints from MACHO and EROS about a decade ago may be summarised as follows:
\begin{equation}
	f( M )
		<
				\begin{cases}
						1
					& ( 6 \times 10^{-8}\,M_{\odot}< M < 30\,M_{\odot} )\, ,\\
						0.1
					& ( 10^{-6}\,M_{\odot}< M < 1\,M_{\odot} )\, ,\\
						0.04
					& ( 10^{-3}\,M_{\odot}< M < 0.1\,M_{\odot} )\,.
				\end{cases}
\end{equation}
Similar limits were obtained by the POINT-AGAPE collaboration, which detected $6$ microlensing events in a survey of the Andromeda galaxy \cite{CalchiNovati:2005cd}. Since then further limits have come from the OGLE experiment. The OGLE-II data \cite{2009MNRAS397, Novati:2009kq, Wyrzykowski:2010bh} yielded somewhat weaker constraints but data from OGLE-III \cite{Wyrzykowski:2010mh} and OGLE-IV \cite{Wyrzykowski:2011tr} gave stronger results for the high mass range:
\begin{equation}
	f( M )
		<
				\begin{cases}
						0.2
					& ( 0.1\,M_{\odot}< M < 20\,M_{\odot} )\, ,\\
						0.09
					& ( 0.4\,M_{\odot}< M < 1\,M_{\odot} )\, ,\\
						0.06
					& ( 0.1\,M_{\odot}< M < 0.4\,M_{\odot} )\,.
				\end{cases}
\end{equation}
We include this limit in Fig.~\ref{fig:large}, and Tab.~\ref{tab:ConstraintSummary} but stress that it depends on some unidentified detections being attributed to self-lensing. Later (comparable) constraints combining EROS and OGLE data were presented in Ref.~\cite{Novati:2013fxa}. Recently Kepler data has improved the limits considerably in the low mass range \cite{Griest:2013esa,Griest:2013aaa}:
\begin{equation}
	f( M )
		<
				0.3 \qquad ( 2 \times 10^{-9}\,M_{\odot}< M < 10^{-7}\,M_{\odot} )
				\, .
\end{equation}
It should be stressed that many papers give microlensing limits on $f( M )$ but it is not easy to combine these limits because they use different confidence levels. Also one must distinguish between limits based on positive detections and null detections. The only positive detection in the high mass range comes from Dong \textit{et al.} \cite{Dong:2007px}.

Early studies of the microlensing of quasars \cite{1994ApJ...424..550D} seemed to exclude all the dark matter being in objects with $10^{-3} M_{\odot} < M < 60 M_{\odot}$. However, this limit does not apply in the $\Lambda$CDM picture and so is not shown in Fig.~\ref{fig:large}. More recent studies of quasar microlensing suggest a limit \cite{2009ApJ...706.1451M}
\begin{equation}
	f( M )
		<
				1
				\quad
				( 10^{-3}\,M_{\odot}< M < 60\,M_{\odot} )
				\, .
\end{equation}
However, this limit might not apply in the $\Lambda$CDM picture, and furthermore the paper states only three data points, so the limit is shown as a dashed line in Fig.~\ref{fig:large}. In this context, Hawkins \cite{1993Natur.366..242H} once claimed evidence for a critical density of jupiter-mass objects from observations of quasar microlensing and associated these with PBHs formed at the quark-hadron transition. However, the status of his observations is no longer clear \cite{Zackrisson:2003wu}, so this is not included in Fig.~\ref{fig:large}. Millilensing of compact radio sources \cite{Wilkinson:2001vv} gives a limit which can be approximated as
\begin{equation}
	f( M )
		<
				\begin{cases}
						( M / 2 \times 10^{4}\, M_{\odot} )^{-2}
					& ( M < 10^{5}\,M_{\odot} )\, , \\
						0.06
					& ( 10^{5}\,M_{\odot}< M < 10^{8}\,M_{\odot} )\, , \\
						(M / 4 \times 10^{8}\,M_{\odot} )^{2}
					& ( M > 10^{8}\,M_{\odot} )\, .
				\end{cases}
\end{equation}
Though weaker than other constraints in this mass range, we include this limit in Fig.~\ref{fig:large} and Tab.~\ref{tab:ConstraintSummary}. The lensing of fast radio bursts could imply strong constraints in the range above $10\, M_{\odot}$ but these are not shown in Fig.~\ref{fig:large}, since they are only potential limits \cite{Munoz:2016tmg}.

\subsection{Dynamical Constraints}
\label{sec:Dynamical-Constraints}

The effects of PBH collisions on astronomical objects{\,---\,}including the Earth \cite{1973Natur.245...88J}{\,---\,}have been a subject of long-standing interest \cite{Carr:1997cn}. For example, Zhilyaev \cite{Zhilyaev:2007rx} has suggested that collisions with stars could produce $\gamma$-ray bursts and Khriplovich \textit{et al.} \cite{Khriplovich:2008er} have examined whether terrestrial collisions could be detected acoustically. Gravitational-wave observatories in space might detect the dynamical effects of PBHs. For example, eLISA could detect PBHs in the mass range $10^{14}$ -- $10^{20}\,{\grm}$ by measuring the gravitational impulse induced by any nearby passing one \cite{Adams:2004pk, Seto:2004zu}. However, we do not show these constraints in Fig.~\ref{fig:large} since they are only potential.

Roncadelli \textit{et al.} \cite{Roncadelli:2009qj} have suggested that halo PBHs could be captured and swallowed by stars in the Galactic disk. The stars would eventually be accreted by the holes, producing a lot of radiation and a population of subsolar black holes which could only be of primordial origin. They argue that every disc star would contain such a black hole if the dark matter were in PBHs smaller than $3 \times 10^{26}$\,g and the following analytic argument \cite{Carr:2009jm} gives the form of the constraint. Since the time-scale on which a star captures a PBH scales as $\tau_{\mathrm{cap}} \propto n_{\mathrm{PBH}}^{-1} \propto M\.f( M )^{-1}$, requiring this to exceed the age of the Galactic disc implies
\begin{equation}
	f
		<
				( M / 3 \times 10^{26}\,\grm )
				\, ,
				\label{neutron}
\end{equation}
which corresponds to a \emph{lower} limit on the mass of objects providing the dark matter. A similar analysis of the collisions of PBHs with main-sequence stars, red-giant cores, white dwarfs and neutron stars by Abramowicz \textit{et al.} \cite{Abramowicz:2008df} suggests that collisions are too rare for $M > 10^{20}\,\grm$ or produce too little power to be detectable for $M < 10^{20}\,$g. However, in a related argument, Capela {\it et al.} have constrained PBHs as dark-matter candidates by considering their capture by white dwarfs \cite{Capela:2012jz} and neutron stars \cite{Capela:2013yf}. The survival of these objects implies a limit which can be approximated as
\begin{equation}
	f( M )
		<
				\frac{M}{4.7\times 10^{24}\,\grm}
				\left(
					1
					-
					\exp
					\left[
						-
						\frac{ M }{ 2.9 \times 10^{23}\,\grm }
					\right]
				\right)^{\!\!-1}
				\qquad
				\left(
					2.5 \times 10^{18} \grm < M < 10^{25}\,\grm
				\right)
				.
\end{equation}
This is similar to Eq.~\eqref{neutron} at the high-mass end, the upper cut-off at $10^{25}$\,g corresponding to the condition $f = 1$. There is also a lower cut-off at $2 \times 10^{18}$\,g because PBHs lighter than this will not have time to consume the neutron stars during the age of the Universe. This argument assumes that there is dark-matter at the centers of globular clusters and is sensitive to the dark-matter density there (taken to be $10^{4}$\,GeV$\,$cm$^{-3}$). Pani \& Loeb \cite{Pani:2014rca} have argued that this excludes PBHs from providing the dark matter throughout the sublunar window, although this has been disputed \cite{Capela:2014qea, Defillon:2014wla}. In fact, the dark-matter density is limited to much lower values than assumed above for particular globular clusters \cite{Ibata:2012eq, Bradford:2011aq}.

Binary star systems with wide separation are vulnerable to disruption from encounters with MACHOs \cite{1985ApJ...290...15B,1987ApJ...312..367W}. Observations of wide binaries in the Galaxy therefore constrain the abundance of halo PBHs. By comparing the results of simulations with observations, Yoo \textit{et al.} \cite{Yoo:2003fr} originally ruled out MACHOs with $M > 43\,M_{\odot}$ from providing the dark matter. However, a careful analysis by Quinn \textit{et al.} \cite{Quinn:2009zg} of the radial velocities of these binaries found that the widest-separation one was spurious, so that the constraint became
\begin{equation}
	f( M )
		<
				\begin{cases}
						( M / 500\,M_{\odot} )^{-1}
					& ( 500\,M_{\odot}< M \lesssim 10^{3}\,M_{\odot} )\, , \\
						0.4
					& ( 10^{3}\,M_{\odot}\lesssim M < 10^{8}\,M_{\odot} )\,.
				\end{cases}
\end{equation}
It flattens off above $10^{3}\,M_{\odot}$ because the encounters are non-impulsive there. Although not shown in Fig.~\ref{fig:large}, more recent studies by Monroy-Rodriguez \& Allen reduce the mass at which $f$ can be $1$ from $500\,M_{\odot}$ to $21$ -- $78\,M_{\odot}$ or even $7$ -- $12\,M_{\odot}$ \cite{Monroy-Rodriguez:2014ula}. The narrow window between the microlensing lower bound and the wide-binary upper bound is therefore shrinking and may even have been eliminated altogether (see Sec.~\ref{sec:ExtendedPBH}).

A variety of dynamical constraints come into play at higher mass scales. These have been studied by Carr and Sakellariadou \cite{Carr:1997cn} and apply providing there is at least one PBH per galactic halo. This corresponds to the condition
\begin{equation}
	f( M )
		>
				( M / M_{\mathrm{halo}} ),
				\quad
	M_{\mathrm{halo}}
		\approx
				3 \times 10^{12}\,M_{\odot}
				\, ,
				\label{incredulity}
\end{equation}
which they term the ``incredulity limit''. An argument similar to the binary disruption one shows that the survival of globular clusters against tidal disruption by passing PBHs gives a limit (not shown in Fig.~\ref{fig:large})
\begin{equation}
	f( M )
		<
				\begin{cases}
						( M / 3 \times 10^{4}\,M_{\odot} )^{-1}
					& ( 3 \times 10^{4}\,M_{\odot} < M < 10^{6}\,M_{\odot} )\, , \\
						0.03
					& ( 10^{6}\,M_{\odot} < M < 10^{11}\,M_{\odot} )\, , \\
						( M / M_{\mathrm{halo}} )
					& ( M > 10^{11}\,M_{\odot} ) \, ,
				\end{cases}
\end{equation}
although this depends sensitively on the mass and the radius of the cluster. The limit flattens off above $ 10^{6}\,M_{\odot}$ because the encounter becomes non-impulsive (\cf~the binary case). The upper limit of $ 3 \times 10^{4}\,M_{\odot}$ on the mass of objects dominating the halo is consistent with the numerical calculations of Moore \cite{1993ApJ...413L..93M}. In a related limit, Brandt \cite{Brandt:2016aco} claims that a mass above $5\, M_{\odot}$ is excluded by the fact that a star cluster near the centre of the dwarf galaxy Eridanus II has not been disrupted by halo objects. His constraint can be written as
\begin{equation}
	f( M )
		\lesssim 	
				\begin{cases}
						( M / 3.7\,M_{\odot} )^{-1} / [ 1.1 - 0.1\ln( M / M_{\odot} ) ]
					& ( M < 10^{3} M_{\odot} ) \, ,\\
						( M / 10^{6} M_{\odot} )
					& ( M > 10^{3} M_{\odot} ) \, ,
				\end{cases}
				\label{eq:eribound}
\end{equation}
where the density of the dark matter at the center of the galaxy is taken to be $0.1\,M_{\odot}\.\mathrm{pc}^{-3}$, the velocity dispersion there is taken to be $5\.{\rm km}\.{\rm s}^{-1}$, and the age of the star cluster is taken to be $3\,{\rm Gyr}$. The second expression in Eq.~\eqref{eq:eribound} was not included in Ref.~\cite{Brandt:2016aco} but is the incredulity limit, corresponding to having one black hole for the dwarf galaxy.

Halo objects will overheat the stars in the Galactic disc unless one has \cite{Carr:1997cn}
\begin{equation}
	f( M )
		<
				\begin{cases}
						( M / 3 \times 10^{6}\,M_{\odot} )^{-1}
					& ( M < 3 \times 10^{9}\,M_{\odot} ) \, , \\
						( M / M_{\mathrm{halo}} )
					& ( M > 3 \times 10^{9}\,M_{\odot} ) \, ,
				\end{cases}
				\label{disc}
\end{equation}
where the lower expression is the incredulity limit. The upper limit of $3 \times 10^{6}\,M_{\odot}$ agrees with the more precise calculations by Lacey and Ostriker \cite{1985ApJ...299..633L}, although they argued that black holes with $2 \times 10^{6}\,M_{\odot}$ could \emph{explain} some features of disc heating. Constraint \eqref{disc} bottoms out at $M \sim 3 \times 10^{9}\,M_{\odot}$ with a value $f \sim 10^{-3}$. Evidence for a similar effect may come from the claim of Totani \cite{Totani:2009af} that elliptical galaxies are puffed up by dark halo objects of $10^{5}\,M_{\odot}$. These disk-heating limits are not shown in Fig.~\ref{fig:large} because they are smaller than other limits in this mass range.

Another limit in this mass range arises because halo objects will be dragged into the nucleus of our own Galaxy by the dynamical friction of the spheroid stars and halo objects themselves (if they have an extended mass function), this leading to excessive nuclear mass unless \cite{Carr:1997cn}
\begin{equation}
	f( M )
		<
				\begin{cases}
						( M / 2 \times 10^{4}\,M_{\odot} )^{-10/7}\,( r_{\mathrm c} / 2\,\mathrm{kpc} )^{2}
					& ( M < 5 \times 10^{5}\,M_{\odot} )\, , \\
						( M / 4 \times 10^{4}\,M_{\odot} )^{-2}\,( r_{\mathrm c} / 2\,\mathrm{kpc} )^{2}
					& ( 5 \times 10^{5}\,M_{\odot}\ll M < 2 \times 10^{6}\,( r_{\mathrm c} / 2\,\mathrm{kpc} )\,M_{\odot} )\, , \\
						( M / 0.1\,M_{\odot} )^{-1/2}
					& ( 2 \times 10^{6}\,(r_{\mathrm c} / 2\,\mathrm{kpc} )\,M_{\odot} < M < 10^{7} M_{\odot} )\, , \\
						( M / M_{\mathrm{halo}} )
					& ( M > 10^{7} M_{\odot} ) \, .
				\end{cases}
\end{equation}
The last expression is the incredulity limit and first three correspond to the drag being dominated by spheroid stars (low $M$), halo objects (high $M$) and some combination of the two (intermediate $M$). The limit bottoms out at $M \sim 10^{7} \, M_{\odot}$ with a value $f \sim 10^{-5}$ but is sensitive to the halo core radius $r_{\mathrm c}$. Also there is a caveat here in that holes drifting into the nucleus might be ejected by the slingshot mechanism if there is already a binary black hole there \cite{Hut:1992iy}. This possibility was explored by Xu and Ostriker \cite{Xu:1994vb}, who obtained an upper limit of $3 \times 10^{6}M_{\odot}$.

Each of these dynamical constraints is subject to certain provisos but it is interesting that they all correspond to an upper limit on the mass of the objects which dominate the halo in the range $500\,-\,2 \times 10^{4}\,M_{\odot}$, the binary-disruption limit being the strongest. This is particularly relevant for constraining models in which the dark matter is postulated to comprise IMBHs. Apart from the Galactic disc and elliptical galaxy heating arguments of Refs.~\cite{1985ApJ...299..633L,Totani:2009af}, it must be stressed that none of these dynamical effects gives {\it positive} evidence for MACHOs. Furthermore, none of them requires the MACHOs to be PBHs. Indeed, they could equally well be clusters of smaller objects \cite{1987ApJ...316...23C,Belotsky:2015psa} or Ultra-Compact Mini-Halos (UCMHs) \cite{Bringmann:2011ut}. This is pertinent in light of the claim by Dokuchaev {\it et al.} \cite{Dokuchaev:2004kr} and Chisholm \cite{Chisholm:2005vm} that PBHs could form in tight clusters, giving a local overdensity well in excess of that provided by the halo concentration alone. It is also important to note that the UCMH constraints on the density perturbations may be stronger than the PBH limits in the higher-mass range \cite{Bringmann:2011ut}. This is relevant if one wants to consider the effect of an extended mass function.

\subsection{Large-Scale Structure Constraints}
\label{sec:Large--Scale-Structure-Constraints}

Sufficiently large PBHs could have important consequences for large-scale structure formation because of the Poisson fluctuations in their number density. This effect was first pointed out by M\'esz\'aros \cite{Meszaros:1975ef} and subsequently studied by various authors \cite{1983ApJ...275..405F, 1977A&A....56..377C, 1983ApJ...268....1C}. In particular, Afshordi \textit{et al.} \cite{Afshordi:2003zb} used observations of the Lyman-$ \alpha $ forest to obtain an upper limit of about $ 10^{4}\,M_{\odot}$ on the mass of any PBHs which provide the dark matter. Although this conclusion was based on numerical simulations, Carr \textit{et al.} \cite{Carr:2009jm} obtained this result analytically and extended it to the case where the PBHs only provide a fraction $ f( M ) $ of the dark matter. Since the Poisson fluctuation in the number of PBHs on a mass-scale $M_{\mathrm{Ly}\alpha} \sim 10^{10}\,M_{\odot}$ grows between the redshift of CDM domination ($z_{\mathrm{eq}} \sim 4000$) and the redshift at which Lyman-$ \alpha $ clouds are observed ($z_{\mathrm{Ly}\alpha} \sim 4$) by a factor $ z_{\mathrm{eq}} / z_{\mathrm{Ly}\alpha} \sim 10^{3} $, the clouds will bind too early unless
\begin{equation}
	f( M )
		<
				\begin{cases}
						 ( M / 10^{4}\,M_{\odot} )^{-1}
						 ( M_{\mathrm{Ly}\alpha} / 10^{10}\,M_{\odot} )
					& ( M < 10^{7} M_{\odot} ) \, , \\
						( M / 10^{10}\,M_{\odot} )
						( M_{\mathrm{Ly}\alpha} / 10^{10}\,M_{\odot} )^{-1}
					& ( M > 10^{7}\,M_{\odot} ) \, .
				\end{cases}
				\label{eq:cluster}
\end{equation}
The lower expression corresponds to having at least one PBH per Lyman-$\alpha$ mass, so the limit bottoms out at $M \sim 10^{7}\,M_{\odot}$ with a value $f \sim 0.001$. The data from SDSS are more extensive \cite{McDonald:2004eu}, so the limiting mass may now be reduced. A similar effect can allow clusters of large PBHs to evolve into the supermassive black holes in galactic nuclei \cite{1984MNRAS.206..801C, Duechting:2004dk, Khlopov:2004sc}; if one replaces $M_{\mathrm{Ly}\alpha}$ with $ 10^{8}\,M_{\odot}$ and $z_{\mathrm{Ly}\alpha}$ with $10$ in the above analysis, the limiting mass in Eq.~\eqref{eq:cluster} is reduced to $ 600\,M_{\odot}$\,.

Recently, Kashlinksy has been prompted by the LIGO observations to consider the effects of the Poisson fluctuations induced by a dark-matter population of $30\,M_{\odot}$ black holes \cite{Kashlinsky:2016sdv}. This can be seen as a special case of the general analysis presented above. However, he adds an interesting new feature to the scenario by suggesting that the black holes might also lead to the cosmic infrared background (CIB) fluctuations detected by the {\it Spitzer/Akari} satellites \cite{Kashlinsky:2014jja, Helgason:2015ema}. This is because the associated Poisson fluctuations would allow more abundant early collapsed halos than in the standard scenario. It has long been appreciated that the CIB and its fluctuations would be a crucial test of any scenario in which the dark matter comprises the black-hole remnants of Population III stars \cite{Bond:1985pc}, but in this case the PBHs are merely triggering high-redshift star formation and not generating the CIB directly. We do not attempt to derive constraints on the PBH scenario from the CIB observations, since many other astrophysical parameters are involved.

\subsection{Accretion Constraints}
\label{sec:Accretion-Constraints}

There are good reasons for believing that PBHs cannot grow very much during the radiation-dominated era. Although a simple Bondi-type argument suggests that they could grow as fast as the horizon \cite{1967SvA....10..602Z}, this does not account for the background cosmological expansion and a fully relativistic calculation shows that such self-similar growth is impossible \cite{Carr:1974nx,1978ApJ...219.1043B,1978ApJ...225..237B}. Consequently there is very little growth during the radiation era. The only exception might be if the Universe were dominated by a ``dark energy'' fluid with $p < - \rho\.c^{2} / 3$, as in the quintessence scenario, since self-similar black-hole solutions do exist in this situation \cite{Harada:2007tj, Maeda:2007tk, Carr:2010wk}. This may support the claim of Bean and Magueijo \cite{Bean:2002kx} that intermediate-mass PBHs might accrete quintessence efficiently enough to evolve into the SMBHs in galactic nuclei.

Even if PBHs cannot accrete appreciably in the radiation-dominated era, massive ones might still do so in the period after decoupling and the Bondi-type analysis \emph{should} then apply. The associated accretion and emission of radiation could have a profound effect on the thermal history of the Universe, as first analysed by Carr \cite{1981MNRAS.194..639C}. This possibility was investigated in more detail by Ricotti \textit{et al.} \cite{Ricotti:2007au}, who studied the effects of such accreting PBHs on the ionisation and temperature evolution of the Universe. The emitted X-rays would produce measurable effects in the spectrum and anisotropies of the CMB. Using FIRAS data to constrain the first and WMAP data to constrain the second, they improve the constraints on $f( M )$ by several orders of magnitude for $ M > 1\,M_{\odot}$\,. The WMAP limit can be approximated as
\begin{equation}
	f( M )
		<
				\begin{cases}
						( M / 30\,M_{\odot} )^{-2}
					& ( 30\,M_{\odot}< M \lesssim 10^{4}\,M_{\odot} )\, ,\\
						10^{-5}
					& ( 10^{4}\,M_{\odot}\lesssim M < 10^{11}\,M_{\odot} )\, ,\\
						M / M_{\ell = 100}
					& ( M > 10^{11}\,M_{\odot} )\, ,\\
				\end{cases}
\end{equation}
where the last expression is not included in Ref.~\cite{Ricotti:2007au} but corresponds to having one PBH on the scale associated with the CMB anisotropies; for $\ell = 100$ modes, this is $M_{\ell = 100} \approx 10^{16} M_{\odot}$. The FIRAS limit can be approximated as
\begin{equation}
	f( M )
		<
				\begin{cases}
						( M / 1\,M_{\odot} )^{-2}
					& (1\,M_{\odot} < M \lesssim 10^{3}\,M_{\odot} )\, , \\
						0.015
					& ( 10^{3}\,M_{\odot}\lesssim M < 10^{14}\,M_{\odot} )\, ,\\
						M / M_{\ell = 100}
					& ( M > 10^{14}\,M_{\odot} )\,. \\
				\end{cases}
\end{equation}
Although these limits appear to exclude $f = 1$ down to masses as low as $1\,M_{\odot}$, they are model-dependent (spherically symmetric Bondi accretion etc.) and therefore not as secure as the dynamical ones. In particular, they depend on the duty-cycle parameter; we assume a smaller value for this than Ref.~\cite{Carr:2009jm}, which is why our limits are somewhat weaker. Mack \textit{et al.} \cite{Mack:2006gz} have considered the growth of large PBHs through the capture of dark-matter halos and suggested that their accretion could give rise to ultra-luminous X-ray sources. The latter possibility has also been explored by Kawaguchi \textit{et al.} \cite{Kawaguchi:2007fz}.

In Ref.~\cite{Eroshenko:2016yve} it is claimed that dark matter will cluster around PBHs from very early times, causing sharp density spikes. These would be observable as bright $\gamma$-ray sources from the annihilation of dark-matter particles in orbit around the PBHs. Very stringent constraints on $f$ are obtained using Fermi-LAT data \cite{Abdo:2010nz} for $M > 10^{-8}\,M_{\odot}$. As this constraint depends on the assumption that the dark-matter density is dominated by WIMPs, we do not include it here. However, such PBH limits must be taken into account if they are to be used to constrain models of inflation.

\begin{center}
	\begin{table}	
		\begin{tabular}{ | l | l | p{5cm} |}
			\hline
			\bf Mass range 								& \bf Constraint
													& \bf Source\\
			\hline
			\hline
			$ M < 10^{18}$g							& $f( M ) < 2 \times 10^{-8} 
														\left(
															\frac{M}{5 \times 10^{14^{^{^{}}}}\!\grm}
														\right)^{\!3 + \epsilon^{^{^{}}}}_{_{_{_{}}}}$
													& extragalactic $\gamma$-ray background\\
			\hline 
			$ 5 \times10^{16} \grm < M < 10^{19}$g		& $f( M ) < 0.1$ & femtolensing of GRB from Fermi\\
			\hline
			$ 2.5 \times 10^{18} \grm < M < 10^{25}$g	& $f( M ) < \frac{M}{4.7 \times 10^{24^{^{^{}}}}\!\grm}
														\bigg(
															1
															-
															\exp
															\left[
																-
																\frac{M}{2.9\times 10^{23^{^{^{}}}}\!\grm}
															\right]
															\mspace{-4mu}
														\bigg)^{\!-1^{^{^{}}}}_{_{_{_{}}}}$
													& neutron-star capture\\
			\hline
			$2 \times 10^{-9} M_{\odot} < M < 10^{-7}\.M_{\odot}$
													& $f( M ) < 0.3
														^{^{^{^{}}}}_{_{_{_{}}}}$
													& microlensing from Kepler\\
			\hline

			$ 10^{-6} M_{\odot} < M < M_{\odot}$			& $f( M ) < 0.1
														^{^{^{^{}}}}_{_{_{_{}}}}$	
													& MACHO and EROS, (OGLE II)\\
			\hline
			$10^{-3} M_{\odot} < M < 0.1 M_{\odot}$			& $f( M ) < 0.04
														^{^{^{^{}}}}_{_{_{_{}}}}$
													& MACHO and EROS, (OGLE II)\\
			\hline
			$0.1 M_{\odot} < M < 0.4 M_{\odot}$				& $f( M ) < 0.06
														^{^{^{^{}}}}_{_{_{_{}}}}$
													& OGLE III and OGLE IV\\
			\hline
			$0.1 M_{\odot} < M < 20 M_{\odot}$				& $f( M ) < 0.2
														^{^{^{^{}}}}_{_{_{_{}}}}$
													& OGLE III and OGLE IV\\
			\hline
			$M > M_{\odot}$ 							& $f( M ) < 3.7\,\frac{M_{\odot}}{M^{^{^{^{}}}}\!}
														\bigg(
															1.1
															+
															0.1
															\ln\!\left[ \frac{M_{\odot}}{M^{^{^{^{}}}}\!} \right]
															\mspace{-4mu}
														\bigg)^{\!-1^{^{^{}}}}_{_{_{_{}}}}$
													& Eridanus II star cluster\\
			\hline
			$500\.M_{\odot} < M < 10^{3} M_{\odot}$			& $f( M ) < \frac{500 M_{\odot}}{M^{^{^{^{}}}}\!}
														^{^{^{^{}}}}_{_{_{_{}}}}$
													& wide-binary stability\\
			\hline
			$10^{3} M_{\odot} < M < 10^{8} M_{\odot}$		& $f( M ) < 0.4^{^{^{^{}}}}_{_{_{_{}}}}$
													& wide-binary stability\\
			\hline
			$ 10\.M_{\odot} < M < \times 10^{4} M_{\odot}$ 	& $f( M ) <
														\Big(
															\frac{M}{10 M_{\odot}^{^{^{^{}}}}\!}
														\Big)^{\!-2^{^{^{}}}}_{_{_{_{}}}}$
													& WMAP3 accretion\\
			\hline
			$M > 10^{4} M_{\odot}$						& $f( M ) < \mathrm{max}
														\bigg[
															10^{-5},
															\left(
																\frac{M}{10^{16^{^{^{}}}}\!M_{\odot}}
															\right)\mspace{-4mu}
														\bigg]
														^{^{^{^{}}}}_{_{_{_{}}}}$
													& WMAP3 accretion\\
			\hline													
			$M > 10^{4} M_{\odot}$						& $f( M ) < {\rm max}\!
														\left[
															\frac{10^{4} M_{\odot}}{M^{^{^{^{}}}}\!}\,,
															\frac{M}{10^{10^{^{^{}}}}\!M_{\odot}}
														\right]
														^{^{^{^{}}}}_{_{_{_{}}}}$
													& Lyman-$\alpha$ clouds\\
			\hline
			$M < 5\times 10^{5} M_{\odot}$					& $f( M ) < 
														\!\left(
															\frac{M}{2\times 10^{4^{^{^{}}}}\!M_{\odot}}
														\right)^{\!-10 / 7^{^{^{}}}}_{_{_{_{}}}}$
													& dynamical friction\\
			\hline
			$5\times 10^{5} M_{\odot} < M < 2 \times 10^{6} M_{\odot}$
													& $f( M ) <
														\!\left(
															\frac{M}{4\times 10^{4^{^{^{}}}}\!M_{\odot}}
														\right)
														^{\!-2^{^{^{}}}}_{_{_{_{}}}}$
													& dynamical friction\\
			\hline
			$M > 2 \times 10^{6} M_{\odot}$ 				& $f( M ) < \mathrm{max}
														\bigg[\!
															\Big(
																\frac{M}{0.1 M_{\odot}^{^{^{^{}}}}\!}
															\Big)^{\!-1/2^{^{^{}}}},
															\frac{M}{3\times 10^{12^{^{^{}}}}\!M_{\odot}}
														\bigg]^{^{^{}}}_{_{_{}}}$
													& dynamical friction\\
			\hline
			$M < 10^{5} M_{\odot}$ 						& $f( M ) <
														\!\left(
															\frac{M}{2 \times 10^{4^{^{^{}}}}\!M_{\odot}}
														\right)^{\!-2^{^{^{}}}}_{_{_{_{}}}}$
													& millilensing of quasars\\	
			\hline
			$10^{5} M_{\odot} < M < 10^{8} M_{\odot}$ 		& $f( M ) < 0.06^{^{^{^{}}}}_{_{_{_{}}}}$
													& millilensing of quasars\\										
			\hline
			$M > 10^{8} M_{\odot}$ 						& $f( M ) <
														\!\left(
															\frac{M}{4 \times 10^{8^{^{^{}}}}\!M_{\odot}}
														\right)^{\!2^{^{^{}}}}_{_{_{_{}}}}$
													& millilensing of quasars\\	
			\hline 
		\end{tabular}
		\caption{Summary of dominant constraints on the fraction of dark matter in PBHs in various mass ranges. 
			These correspond to or are special cases of the constraints in Fig.~\ref{fig:large}; 
			see main text for details. Only limits stronger than $f( M ) < 1$ are listed.}
		\vs{-3mm}
		\label{tab:ConstraintSummary}
	\end{table}
\end{center}

\section{Confronting extended PBH mass functions with constraints}
\label{sec:ExtendedPBH}

\noindent
The constraints discussed above are usually applied on the assumption that the PBH mass function is nearly monochromatic (\ie~with a width $\Delta M \sim M$). However, this is unrealistic and we discussed in Sec.~\ref{sec:RealisticMassFunctions} scenarios in which one would expect the mass function to be extended. In the context of the dark-matter problem, this is a two-edged sword. On the one hand, it means that the {\it total} PBH density may suffice to explain the dark matter, even if the density in any particular mass band is small and within the observational bounds discussed in Sec.~\ref{sec:Constraints}. On the other hand, even if PBHs can provide all the dark matter at some mass-scale without violating the constraints there, the extended mass function may still violate the constraints at some other scale (even if $f$ is low there). In view of the numerous constraints in Fig.~\ref{fig:large}, this problem is particularly pertinent if the mass function extends over many decades.

In general, a detailed assessment of these two ``edges of the sword'' requires a knowledge of the forms of both the expected PBH mass fraction, $f_{\rm exp}( M )$, and the maximum fraction allowed by the constraint, $f_{\rm max}( M )$. The procedure is non-trivial even when the forms of these functions can be expressed analytically. In particular, one cannot just plot $f_{\rm exp}( M )$ for a given model in Fig.~\ref{fig:large} and infer that the model is allowed because it does not intersect $f_{\rm max}( M )$, even though this procedure is sometimes used in the literature \cite{Clesse:2015wea}. For example, if the constraint has the ``flat'' form $f( M ) < a$ over the range $M_{\rm min}$ to $M_{\rm max}$, then the total fraction of PBHs in this mass range cannot exceed $a$, so one must integrate the function $ \d f_{\rm exp} / \d M$ from $M_{\rm min}$ to $M_{\rm max}$. A more general constraint can be treated as a sequence of flat constraints by breaking it up into narrow mass bins. One starts near the minimum of the constraint and defines a bin from $M_{1}$ to $M_{2}$ around this. The constraint in this range can be approximated by $f < q_{\rm max}$, where $q_{\rm max}$ is the maximum value of $f_{\rm max}$ between these masses, being chosen so that $q_{\rm max}$ is very close to the minimum of $f_{\rm max}$ and comparable to the integrated mass function $f_{\rm exp}$ in this region. We then move to the next bin, $M_{3} \leq M_{1}$ to $M_{4} \geq M_{2}$, and repeat the process. In realistic cases we need only move one of the boundaries, as the mass function will have a single feature which fits on one side of the constraint. We will demonstrate this methodology for particular mass ranges in the subsections below.

When there are different constraints in the range for which $f_{\rm exp}( M )$ is non-zero, this procedure must be repeated for each one. However, the interpretation of intersecting constraints is subtle for an extended mass function. If two constraints meet at $M = M_{\rm meet}$, there can be a fraction $f_{\rm meet} \equiv f( M_{\rm meet} )$ in PBHs both below and above $M_{\rm meet}$, making the combined constraint $f < 2\.f_{\rm meet}$ in the appropriate range, unless some other (stronger) constraint applies there. Hence for a limit which is independent of the PBH formation mechanism, all the constraints in the relevant mass range must sum up to $f < 1$. As discussed below, unless one invokes an extended mass function with multiple maxima, all mass ranges could be excluded in principle. However, there are currently still windows where the model parameters are insufficiently known to exclude PBHs from providing all the dark matter.

\begin{figure}
	\includegraphics[scale=1,angle=0]{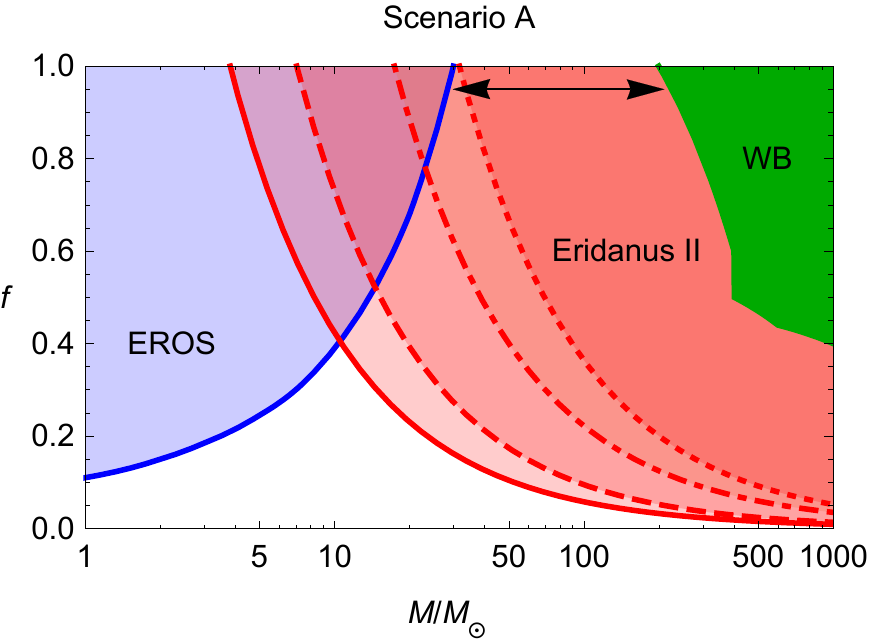}
	\includegraphics[scale=1,angle=0]{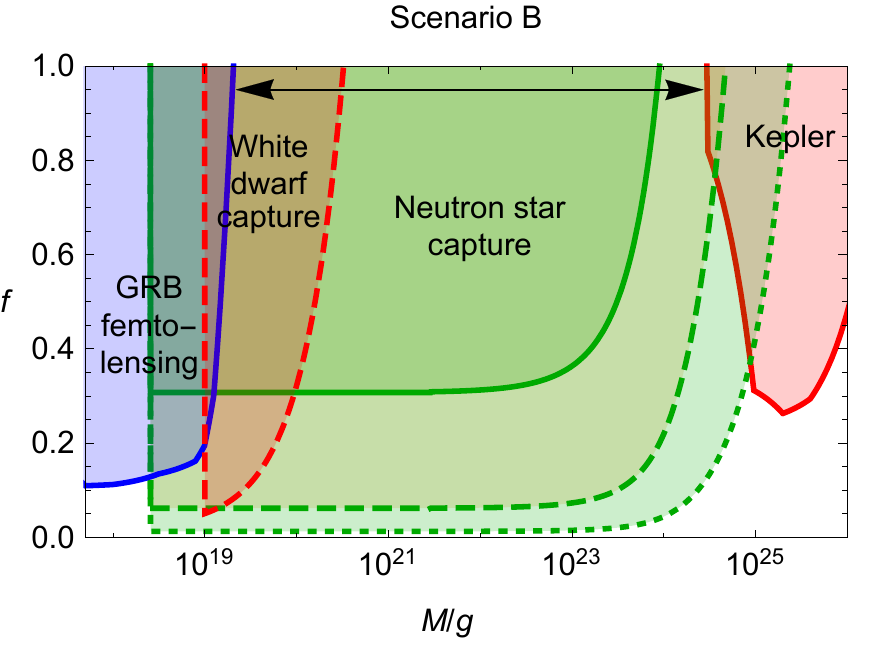}\\[3mm]
	\includegraphics[scale=1,angle=0]{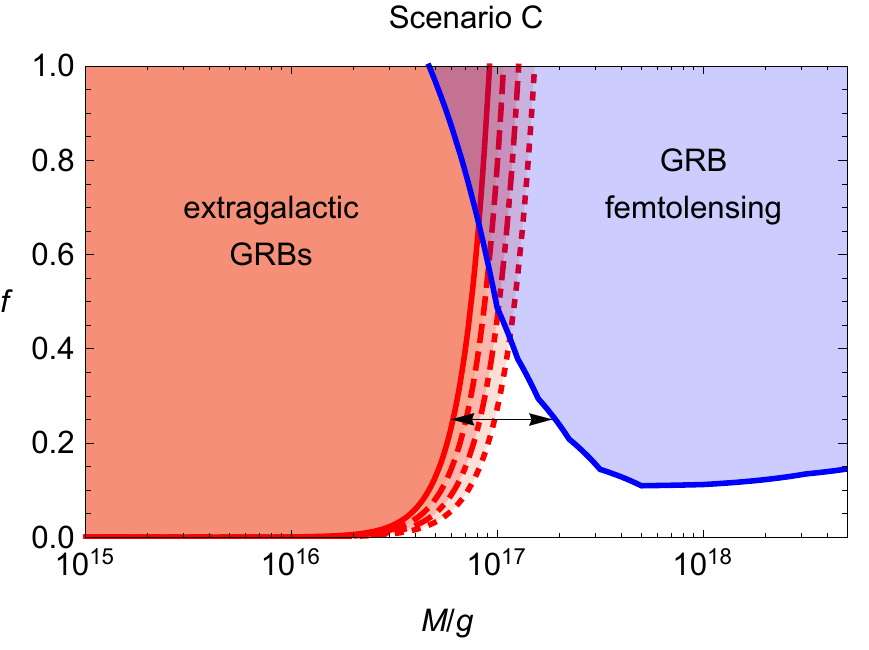}
	\includegraphics[scale=1,angle=0]{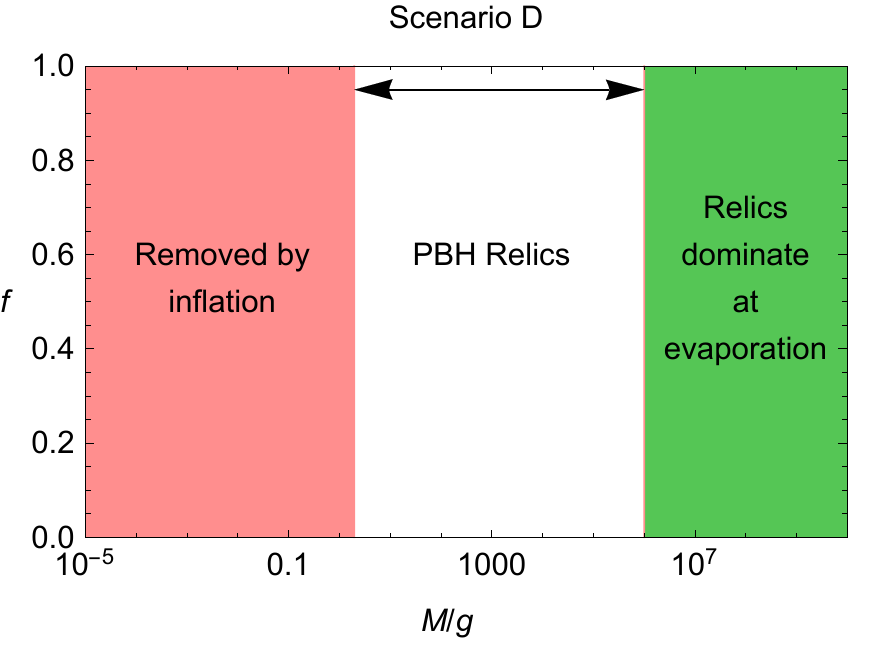}
	\caption{Four windows in which PBHs could conceivably provide the dark-matter density. 
		{\it Upper left panel$\mspace{1.5mu}$}:~(A) Intermediate-mass black holes.
		The constraints in this mass range are EROS and MACHO microlensing bounds \cite{Tisserand:2006zx} (in blue), 
		dynamical constraints (in red) from the life-time of the central star cluster in the Eridanus II dwarf galaxy 
		\cite{Brandt:2016aco}, as well as dynamical constraints (in green) from the existence of wide-binary star systems 
		\cite{Quinn:2009zg}. 
		{\it Upper right panel$\mspace{1.5mu}$}:~(B) Sublunar black holes; In this case the constraints (in blue) are again
		the femtolensing of GRBs from \cite{Barnacka:2012bm}, while the limits from neutron-star capture (in green) 
		are taken from \cite{Capela:2013yf}. 
		The red-shaded region to the right-hand side of the plot denotes microlensing constraints from the Kepler survey 
		\cite{Griest:2013aaa}, , while the red-shaded region to the plot's left-hand side shows constraints from white-dwarf 
		explosions \cite{2015PhRvD..92f3007G}.
		{\it Lower left panel$\mspace{1.5mu}$}:~(C) Subatomic black holes. The constraints here (red-shaded region) stem 
		from non-detections of extragalactic $\gamma$-rays that would be observable from the evaporation of PBHs 
		of these masses \cite{Carr:2009jm, Carr:2016hva}, and (in blue) femtolensing of $\gamma$-ray bursts (GRBs) 
		taken from Fermi data \cite{Barnacka:2012bm}.
		{\it Lower right panel$\mspace{1.5mu}$}:~(D) Planck-mass relics from PBH evaporations. 
		This shows the mass range of the {\it initial} PBHs if they derive from inflation \cite{Carr:1994ar} 
		but there are no observational constraints on such relics. 
		Details on all these regimes and the meaning of the constraints can be found in the subsections on 
		the respective scenarios.}
	\label{fig:windows}
\end{figure}

The discussion in Sec.~\ref{sec:Constraints} shows that there are three such windows: (A) black holes in the intermediate-mass range $1\,M_{\odot} < M < 10^{3}\,M_{\odot}$ between the microlensing and wide-binary limits; (B) sublunar black holes in the range $10^{20}\,\grm < M < 10^{24}\,\grm$ between the femtolensing and Kepler microlensing limits; (C) subatomic-size black holes in the range $10^{16}\,\rm{g} < M < 10^{17}\,\rm{g}$ between the $\gamma$-ray background and femtolensing limits. There is also a fourth window: (D) Planck-mass relics of Hawking evaporation in the range around $M \sim 10^{-5}\,$g. The main constraints in each of these mass windows are indicated in more detail in Fig.~\ref{fig:windows} but it should be stressed that the windows have a different status. (A) is topical because of the recent LIGO results. (B) may be excluded by the neutron star and white-dwarf limits, although this has been disputed. (C) is perhaps implausible because the range is so narrow, although this possibility is stressed in Ref.~\cite{Carr:2009jm}. (D) is essentially untestable because the relics are too small ($10^{-33}\,$cm) to be detected non-gravitationally. It has been suggested that PBHs in window (A) could naturally arise in various inflationary scenarios \cite{Frampton:2010sw, Bugaev:2011qt, Clesse:2015wea, Kawasaki:2015ppx} but this applies equally for the other windows since the mass-scale is essentially arbitrary.

We now discuss each of the mass windows in turn. For the largest one (A), we will present our analysis in some detail in order to demonstrate the methodology. For the next two mass windows (B and C), we have performed a similar analysis but just state the main results. Finally, the Planck-mass relic scenario (D) is discussed, although there is only the trivial constraint $f < 1$ in this mass range. We stress that we are not making definite conclusions about the viability of PBH dark matter in any particular range. We are merely considering how conclusions can be drawn from certain observational claims in the literature, which may or may not be justified.

\subsection*{Scenario A -- Intermediate-mass Black Holes}

\begin{figure}
	\centering
	\includegraphics[scale=1,angle=0]{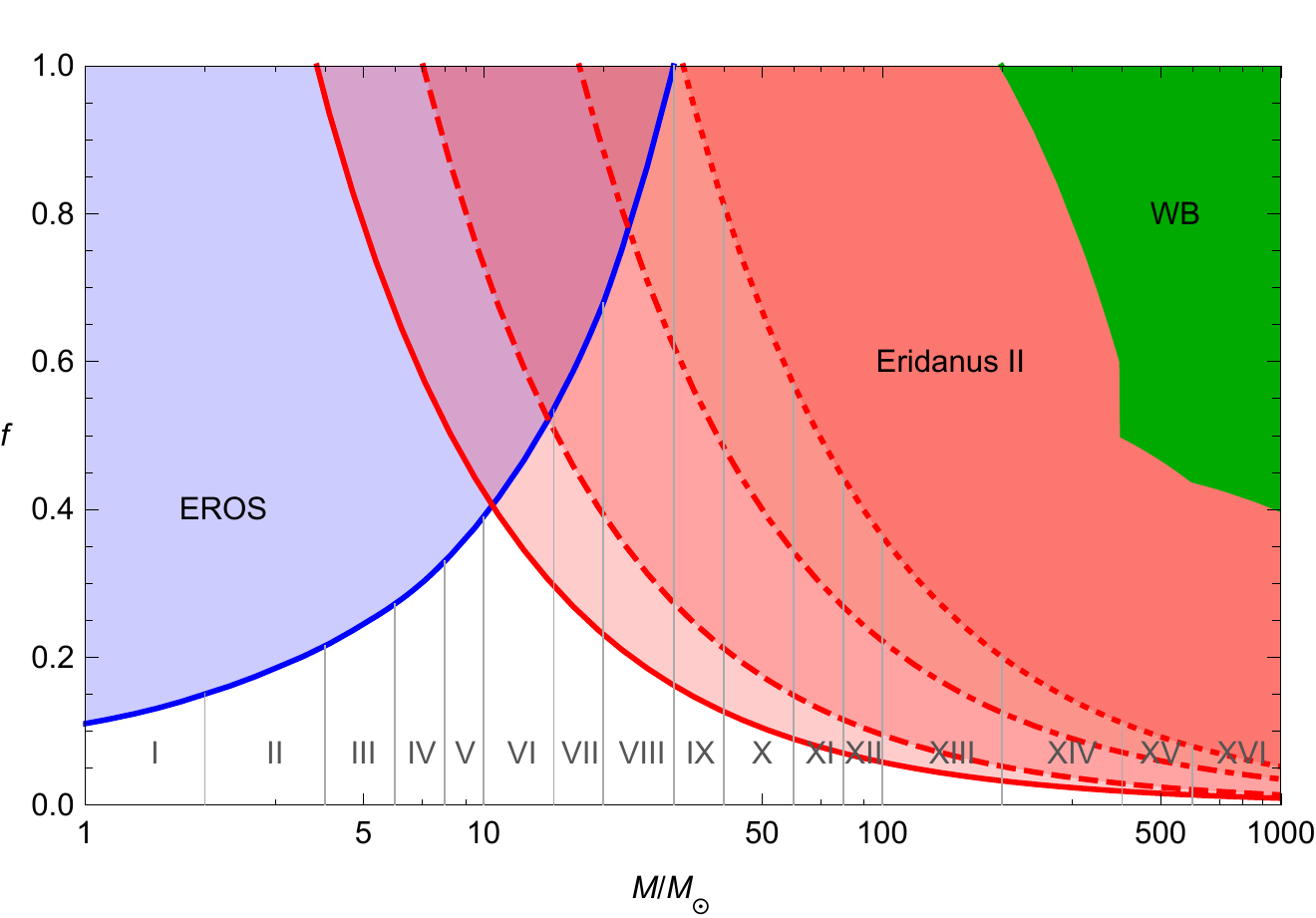}
	\caption{Constraints on the dark-matter fraction of primordial black holes in the intermediate-mass range 
		$M_{\odot} < M < 10^{3}\,M_{\odot}$. Excluded regions are shaded. EROS constraints are taken from 
		Ref.~\cite{Tisserand:2006zx} and are depicted in blue. 
		Wide-binary (WB) constraints \cite{Chaname:2003fn, Yoo:2003fr}
		correspond to the green region in the plot. The latest constraints from the survival of the star cluster near 
		the core of Eridanus II \cite{Brandt:2016aco} are shown in the red-shaded areas. 
		For all red curves we assume 
		a cluster age of $3\,{\rm Gyr}$. The various constraints are due to different 
		choices of values for the velocity dispersion
		$\sigma$ and $\rho$, the dark-matter density in the center of the galaxy. Specifically, we chose 
		$( \sigma, \rho ) = ( 5\,{\rm km}\.{\rm s}^{-1}, 0.1\,M_{\odot}\.{\rm pc}^{-3} )$ (red solid), 
		$( \sigma, \rho ) = ( 10\,{\rm km}\.{\rm s}^{-1}, 0.1\,M_{\odot}\.{\rm pc}^{-3} )$ (red dashed), 
		$( \sigma, \rho ) = ( 5\,{\rm km}\.{\rm s}^{-1}, 0.01\,M_{\odot}\.{\rm pc}^{-3} )$ (red dot-dashed), 
		and $( \sigma, \rho ) = ( 10\,{\rm km}\.{\rm s}^{-1}, 0.01\,M_{\odot}\.{\rm pc}^{-3} )$ (red dotted).}
	\label{fig:constraints-Intermediate}
\end{figure}

The constraints in the intermediate-mass range are shown in Fig.~\ref{fig:windows}A and in more detail in Fig.~\ref{fig:constraints-Intermediate}. We include the Eridanus II limits \cite{Brandt:2016aco} but not the CMB limits \cite{Ricotti:2007au}, since the validity of these has been disputed \cite{Bird:2016dcv} and they anyway depend upon uncertain astrophysical parameters.\footnote{During the preparation of this manuscript, Ref.~\cite{Brandt:2016aco} was updated to include constraints from ultra-faint dwarfs and this may exclude all the dark matter being in PBHs in the intermediate-mass window. A recent analysis by Green~\cite{Green:2016xgy}, using the constraints in the second version of Ref.~\cite{Brandt:2016aco} and slightly different choice of parameters, also suggests this window is excluded. She further claims that there is an error in our methodology but we would argue that this still provides a good approximation for most constraints, even though her method may be more accurate in principle. This relates to the width of the mass bins used in our analysis. Compared to Green's top-hat analysis, our methodology will indeed underestimate the constraints if a small number of bins are employed, so the number must be chosen carefully.}
We omit the microlensing-estimates from Ref.\cite{2009ApJ...706.1451M} as they provide only one point in the relevant mass interval. We als omit the most recent OGLE constraints \cite{2009MNRAS397, Novati:2009kq, Wyrzykowski:2010bh, Wyrzykowski:2010mh, Wyrzykowski:2011tr}, as did the analysis in Ref.~\cite{Brandt:2016aco}, the limit from the lensing of fast radio bursts \cite{Munoz:2016tmg} and the latest wide-binary constraints \cite{Monroy-Rodriguez:2014ula} because these cover the same mass range as the Eridanus II limits. Since the latter are very stringent, we need a more precise expression than Eq.~\eqref{eq:eribound} and care must be taken when considering the associated parameters. The constraint can be written as \cite{Brandt:2016aco}
\begin{align}
	f( M )
		&\lesssim
								0.5\,
									\left(
										1
										+
										\frac{0.046 M_{\odot}{\rm pc}^{-3}}{\rho}
									\right)
									\left(
										\frac{10 M_{\odot}}{M}
									\right)
									\left(
										\frac{\sigma}{10\,{\rm km}\.{\rm s}^{-1}}
									\right)
									/ \left(
									1
									+
									0.1
									\ln
									\left[
										\frac{10 M_{\odot}}{M}
										\left(
											\frac{\sigma}{10\,{\rm km}\.{\rm s}^{-1}}
										\right)^{2}
									\right] \right)
									,
									\label{eq:eribound2}
\end{align}
where $\rho$ is the density and $\sigma$ is the velocity dispersion of the dark matter at the center of the galaxy. This reduces to Eq.~\eqref{eq:eribound} for $\rho = 0.1\,M_{\odot}\.{\rm pc}^{-3}$ (a reasonable upper limit) and $\sigma = 5 \,{\rm km}\.{\rm s}^{-1}$. Equations~\eqref{eq:eribound} and \eqref{eq:eribound2} assume an age of $3\,{\rm Gyr}$ for the star cluster. However, it could be as high as $12\,{\rm Gyr}$ \cite{2016arXiv160408590C}, in which case these equations must be modified and yield tighter constraints \cite{Brandt:2016aco}. 

As can be seen from Fig.~\ref{fig:constraints-Intermediate}, the least restrictive Eridanus II constraint, corresponding to $\rho = 0.01\,M_{\odot}\.{\rm pc}^{-3}$ and $\sigma = 10\,{\rm km}\.{\rm s}^{-1}$, admits a monochromatic function containing all the dark matter at $M \sim 30\,M_{\odot}$ of the kind displayed in the left panel of Fig.~\ref{fig:Critical-collapse}. As observations of the dwarf galaxy and wide binaries improve, this gap may be filled and even the present ones shrink it according to Ref.~\cite{Monroy-Rodriguez:2014ula}. However, a monochromatic mass function is not very physical. A model-independent way of assessing the more realistic extended-mass-function case is to consider where the different constraints cross. For $\rho = 0.1\,M_{\odot}\.{\rm pc}^{-3}$, $\sigma = 5\,{\rm km}\.{\rm s}^{-1}$ (red solid curve), which is also the line chosen in Ref.~\cite{Brandt:2016aco}, the Eridanus II and microlensing constraints cross at $M \sim 10\,M_{\odot}$ and $f \approx 0.4$. This means that $40\%$ of the dark matter can be contained in PBHs with $M < 10\,M_{\odot}$, thereby evading the microlensing bounds, and another $40\%$ in PBHs with $M > 10\,M_{\odot}$, thereby evading the Eridanus II constraints. Hence the Eridanus II and microlensing constraints together exclude PBHs from having more than $~80 \%$ of the dark matter in this intermediate-mass range. The slightly less restrictive Eridanus II constraint with $\rho = 0.1\,M_{\odot}\.{\rm pc}^{-3}$, $\sigma = 10\,{\rm km}\.{\rm s}^{-1}$ (red dashed line) crosses the microlensing constraints at $M \sim 20\,M_{\odot}$ and $f \approx 0.5$, marginally allowing the dark matter to be in PBHs in this range. However, in this case the extended mass function has to be perfectly tuned to fit beneath the bounds, which is unlikely. On the other hand, for $\rho = 0.01\,M_{\odot}\.{\rm pc}^{-3}$, $\sigma = 5\,{\rm km}\.{\rm s}^{-1}$ (red dot-dashed curve) and $\rho = 0.01\,M_{\odot}\.{\rm pc}^{-3}$, $\sigma = 10\,{\rm km}\.{\rm s}^{-1}$ (red dotted line), one could certainly envisage a mass function which provides all the dark matter.

From these and other constraints, an extended mass function contributing an equal density at all mass scales can also be excluded. Even without including the Eridanus II constraints, if such a function extends to the range of microlensing observations, the most restrictive range for these indicates that the mass function cannot make up more than $4 \%$ of the dark matter over the two orders of magnitude from $10^{-3}\,M_{\odot}$ to $0.1\,M_{\odot}$. To get the total dark matter in PBHs, one would then need $50$ orders of magnitude, whereas the widest possible range, ignoring all other observations, would be from $10^{-15}\,M_{\odot}$ to $10^{17}\,M_{\odot}$, which is only $32$ orders of magnitude. In practice, the upper limit may be considerably tighter for both observational and theoretical reasons. Hence some bumpy feature is needed to provide the dark matter without violating the constraints. If the strongest Eridanus II constraints is taken seriously, this bumpy feature cannot be confined to the $30\,M_{\odot}$ region. Instead, the bump must either be located at a lower mass or{\,---\,}if restrictive bounds from neutron-star capture \cite{Capela:2013yf} and star formation \cite{Capela:2012jz} can be trusted{\,---\,}a (camel-like) feature with at least two bumps in the appropriate regions might be necessary to put all the dark matter in PBHs.

\begin{figure}
	\includegraphics[scale=1,angle=0]{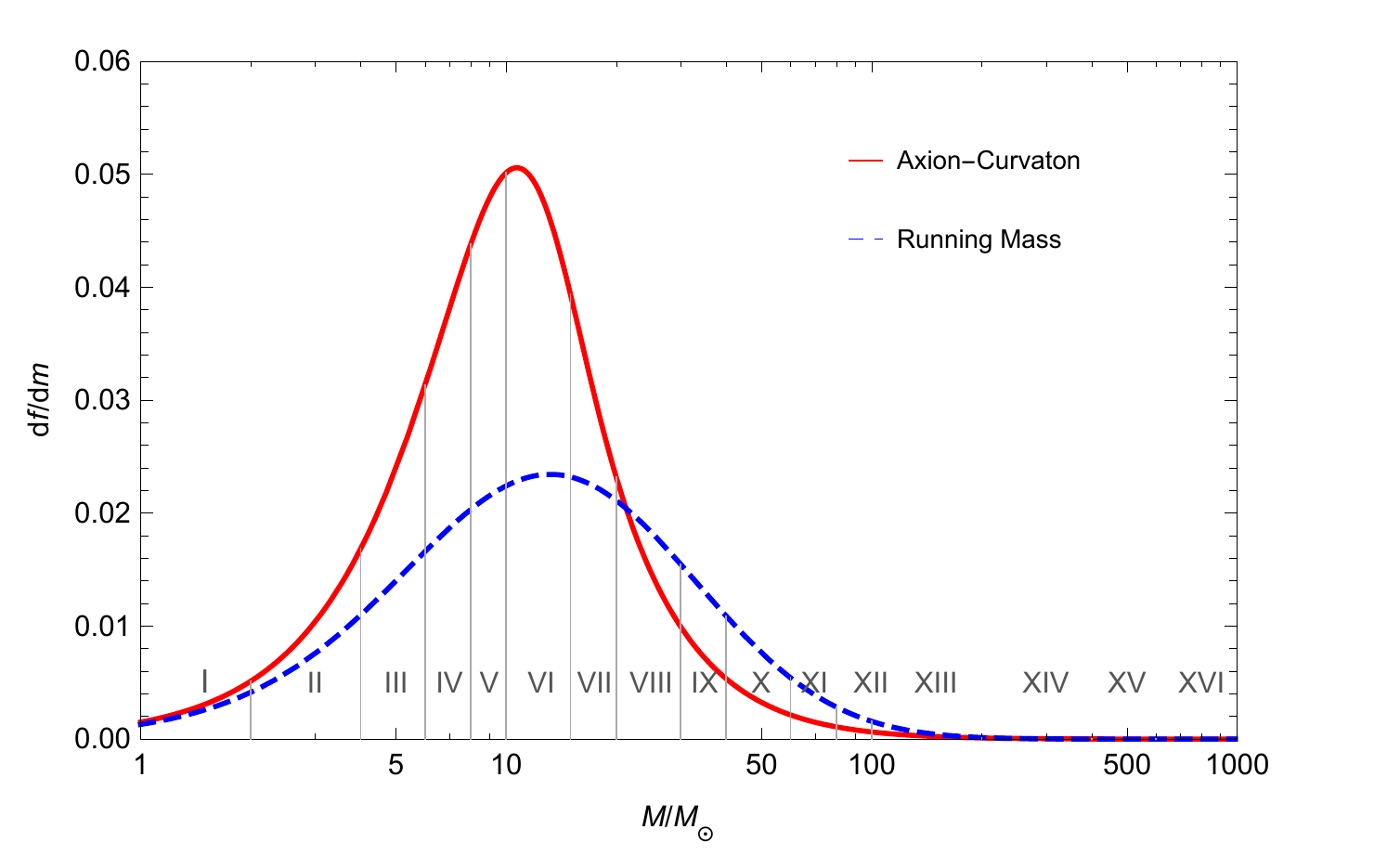}
	\caption{The differential dark-matter fraction $\d f / \d M$ in the intermediate-mass range 
		$M_{\odot} < M < 10^{3}\,M_{\odot}$ 
		for the axion-like curvaton model (red, solid) as well as for running-mass inflation (blue, dashed). 
		The parameter choices are 
		$P_{\zeta}( k_{\frm} ) = 3.08 \times 10^{-3}$, $M_{\rm min} = 6 \times 10^{-7}\,M_{\frm}$, $\lambda = 1.2$ 
		for the axion curvaton model
		(see Sec.~\ref{sec:Axion--Curvaton-Inflation}), and $a = 0.011$, $b = 0.0245$, and $c = -\.0.00304345$ 
		for the running mass model
		(see Sec.~\ref{sec:Running--Mass-Inflation}). These choices are made in order to yield a dark-matter fraction 
		of $1$, as well as to be most compatible with the constraints in the intermediate-mass window 
		(\cf~Fig.~\ref{fig:constraints-Intermediate}). Critical collapse, with $\delta_{c} = 0.45$, $k = 3.3$, 
		and $\gamma = 0.36$, has been applied to obtain both of the curves. 
		Featured are also mass bins/ranges I through XVI used to demonstrate the comparison with constraints 
		as described in the text. The same mass bins can also be seen in Fig.~\ref{fig:constraints-Intermediate} 
		showing the constraints.}
	\label{fig:beta-Eq-comparison}
\end{figure}

To compare the constraints with more realistic models, a more sophisticated approach is needed. We demonstrate this by considering the initial mass functions for the axion-like curvaton and running-mass models shown in Fig.~\ref{fig:beta-Eq-comparison}. We divide the mass range into bins. Starting with the constraints from EROS, one integrates the mass function in the lowest bin (here called I) to obtain the fraction $f( M )$ in this bin. Then this number is compared to the bound on $f$ at the upper end of the bin coming from microlensing studies (\ie~the EROS limit at the right of bin I in Fig.~\ref{fig:constraints-Intermediate}). One then integrates for bins I and II, comparing this to the limit from the upper end of bin II. One then repeats this procedure, running through all the bins. (In this case, there is no reason to go beyond bin VIII.) Similarly, for the Eridanus II constraint, one starts with the largest mass bin (XVI) and integrates to find the dark-matter fraction in this bin. This result is then compared to the constraint at the low end of the bin (the Eridanus II bound at the intersect of bins XV and XVI). Next one combines the integrals from the top two bins (XV and XVI) and compares this to the Eridanus II bound on the intersection between XIV and XV etc.

Using this technique, we have shown that both the axion-like curvaton model and running-mass model featured in Fig.~\ref{fig:beta-Eq-comparison} evade the bounds for the Eridanus II parameters of $\rho = 0.01 - 0.03\,M_{\odot}\.{\rm pc}^{-3}$, $\sigma = 5\,{\rm km}\.{\rm s}^{-1}$ and $\rho = 0.01 - 0.03\,M_{\odot}\.{\rm pc}^{-3}$, $\sigma = 10\,{\rm km}\.{\rm s}^{-1}$, but are ruled out for $\rho = 0.1\,M_{\odot}\.{\rm pc}^{-3}$, $\sigma = 5\,{\rm km}\.{\rm s}^{-1}$ and $\rho = 0.1\,M_{\odot}\.{\rm pc}^{-3}$, $\sigma = 10\,{\rm km}\.{\rm s}^{-1}$. For the Eridanus II parameters with room for a monochromatic mass function, $\sigma = 10\,{\rm km}\.{\rm s}^{-1}$ and $\rho = 0.01\,M_{\odot}\.{\rm pc}^{-3}$, we checked whether the critically-collapsed monochromatic mass function is compatible with the bounds. It turns out that this not only evades the constraints but is also compatible with the Eridanus II parameter values of $\sigma = 10\,{\rm km}\.{\rm s}^{-1}$ and $\rho = 0.03\,M_{\odot}\.{\rm pc}^{-3}$, for which a monochromatic mass function (without critical collapse) is ruled out. Finally, this technique shows the correct way to confront any properly obtained PBH extended mass function with the observational constraints, regardless of how much of the dark matter is in PBHs. If appropriate care is taken to include all influences discussed in Sec.~\ref{sec:RealisticMassFunctions}, this scheme can be used to constrain the inflationary potential and other PBH-producing scenarios. Note that the methodology described here is independent of the particular constraint or model involved. The same procedure both for getting model-independent constraints and confronting models with these constraints can be used regardless the particular constraint or model.\\[-7mm]

\subsection*{Scenario B -- Sublunar-mass Black Holes}

Figure~\ref{fig:windows}B shows the constraints from GRB femtolensing \cite{Barnacka:2012bm} and Kepler microlensing \cite{Griest:2013aaa}. The neutron-star (NS) capture constraint \cite{Capela:2013yf} is also shown but this depends on the assumption that there is dark matter in globular clusters, which is uncertain. If the NS constraint is omitted, $f = 1$ at $10^{20}$g and $4 \times 10^{24}\,$g, respectively, so there are over three decades of mass in which PBHs could provide the dark matter. In this case, it is clear that both monochromatic and extended mass functions can allow $f = 1$. However, the inclusion of the NS constraint \cite{Capela:2013yf} could dramatically alter this situation, depending on the (very uncertain) dark-matter density $\rho_{\rm DM}$ in the core of globular clusters.

The three lines in Fig.~\ref{fig:windows}B correspond to $\rho_{\rm DM} = 4 \times 10^{2}\,{\rm GeV}\.{\rm cm}^{-3}$ (solid), $\rho_{\rm DM} = 2 \times 10^{3}\,{\rm GeV}\.{\rm cm}^{-3}$ (broken) and $\rho_{\rm DM} = 10^{4}\,{\rm GeV}\.{\rm cm}^{-3}$ (dotted). In all of these cases, PBHs are excluded from providing all the dark matter at the lower mass end, where the NS and femtolensing bounds meet. So if the NS bounds are believed, only the window at the upper end is allowed. For the highest dark-matter density, $\rho_{\rm DM} = 10^{4}\,{\rm GeV}\.{\rm cm}^{-3}$, the NS constraint intersects the Kepler constraint at $f_{\rm meet} \approx 0.35$, so PBHs cannot provide all the dark matter, whatever the shape of the mass function. For $\rho_{\rm DM} = 2 \times 10^{3}\,{\rm GeV}\.{\rm cm}^{-3}$ which is suggested by some numerical models \cite{Capela:2013yf}, the constraints cross at $f_{\rm meet} \approx 0.75$, so a monochromatic mass function cannot give all the dark matter but an extended one could allow a fraction $2\.f_{\rm meet} \approx 1.5$. Indeed, if one applies critical collapse to an initially monochromatic mass function, one evades the constraints in the mass range $1$ -- $2 \times 10^{24}\,$g. For the lowest dark-matter density shown, $\rho_{\rm DM} = 4 \times 10^{2}\,{\rm GeV}\.{\rm cm}^{-3}$, even a monochromatic mass function is allowed, so it is important to stress that the density is known to be as low as $1\,{\rm GeV}\.{\rm cm}^{-3}$ for some globular clusters \cite{Ibata:2012eq, Bradford:2011aq}. Indeed, according to the scenario of Ref.~\cite{Capela:2013yf}, dark-matter densities below $120\,{\rm GeV}\.{\rm cm}^{-3}$ always lead to constraints above $f = 1$.

For an extended mass function we have performed a similar analysis to that for the intermediate-mass case. For the axion-curvaton model with $P_{\zeta}( k_{\frm} ) = 5.7442 \times 10^{- 3}$, $M_{\rm min} = 10^{- 8}\.M_{\frm}$, $\lambda = 1.13$, the NS constraint is satisfied for the two lowest values of the dark-matter density, so all the dark matter could be in PBHs. For the running-mass model, the parameter choice $a = 0.011$, $b = 0.00633$ and $c = -\,0.0005399$ yields the best fit. However, this model could evade the NS and microlensing constraints only for the lowest dark-matter density, $\rho_{\rm DM} = 4 \times 10^{2}\,{\rm GeV}\.{\rm cm}^{-3}$. For both models, a critical collapse mass function with $\delta_{c} = 0.45$, $k = 3.3$, and $\gamma = 0.36$ was assumed. This demonstrates the ``two-edged sword'' feature: while the highly peaked axion-curvaton model can fit between the tighter constraints, the more extended running-mass mass model cannot. Of course, a generic running-mass model with an arbitrary number of parameters could be equally steep and would then also evade the bounds. Similarly, while the NS constraint with $\rho_{\rm DM} = 2 \times 10^{3}\,{\rm GeV}\.{\rm cm}^{-3}$ and microlensing constraint exclude a monochromatic mass function from providing all the dark matter, a mass function obtained from the critical collapse of a monochromatic overdensity of the kind exemplified in Fig.~\ref{fig:Critical-collapse} could do so.

\subsection*{Scenario C -- Subatomic-sized Black Holes}

The PBHs considered in this subsection have a radius in the range $10^{-12}$ to $10^{-10}\,$cm ($10$ to $100$ Fermi), so we describe these as ``subatomic''. Figure \ref{fig:windows}C shows the constraint \eqref{photon2} from the extragalactic $\gamma$-ray background, with the dotted curves corresponding to values of $\epsilon$ between $0.1$ and $0.4$, and the constraint \eqref{femto} from the femtolensing of GRBs \cite{Barnacka:2012bm}.\footnote{There is considerable discrepancy between the constraint presented in the published and arXiv version of this paper; we use the former.} These limits hit $f = 1$ at around $M = 10^{17}\,$g and $M = 10^{16.5}\,$g, respectively, so there is no range where $f = 1$ is possible. This means that a monochromatic mass function cannot provide all the dark matter in this case. We see that $f_{\rm meet} \approx 0.4$ when the slope of the observed background is taken to be $\epsilon = 0.1$, so PBHs cannot make up more than $80\%$ of the dark matter in the mass range from $10^{15}$ -- $10^{19}\,$g. With $\epsilon = 0.2$, $f_{\rm meet}$ is just below $0.5$, so PBHs dark matter is marginally excluded in this range. However, with $\epsilon = 0.4\,(0.3)$, we have $f_{\rm meet} \approx 0.55 \, (0.65)$, so a suitably shaped extended mass function could still provide the dark matter. Several authors have argued that this would permit PBHs to explain both the $\gamma$-ray background and the dark matter \cite{Yokoyama:1998xd,Green:1999xm}, with Belotsky \textit{et al.} suggesting that an extended PBH mass function could simultaneously explain the dark matter, the reionisation of the Universe and the annihilation-line radiation from the Galactic centre \cite{Belotsky:2015rxx}.

Performing a more detailed analysis, we find that neither an axion-curvaton nor running-mass model can provide all the dark matter in this window without violating the bounds. In fact, not even a mass function resulting from the critical collapse of a monochromatic overdensity feature allows this. This is mainly due to the steepness of the extragalactic $\gamma$-ray background constraint at the low-mass end of this window. In principle, one could envisage other effects (\eg~accretion or mergers) creating a mass function with a different shape. However, it is extremely unlikely that these effects would conspire to allow the mass function to fit within the bounds. Therefore PBHs are probably excluded from providing all the dark matter in this region, although they could still provide some of it. Ref.~\cite{Carr:2016hva} claims that such PBHs {\it could} provide the dark matter but does not consider the femtolenisng limit.

\subsection*{Scenario D -- Planck-Mass Relics}

If PBH evaporations leave stable Planck-mass relics, these might also contribute to the dark matter. This was first pointed out by MacGibbon \cite{MacGibbon:1987my} and has subsequently been explored in the context of inflationary scenarios by many authors \cite{Barrow:1992hq, Carr:1994ar, Green:1997sz, Alexeyev:2002tg, Chen:2002tu, Barrau:2003xp, Chen:2004ft, Nozari:2005ah}. If the relics have a mass $ \kappa\,M_{\rm Pl} $\,, where $M_{\rm Pl}$ is the Planck mass and $\kappa = \Ocal( 1 )$, and if reheating occurs at a temperature $T_{\Rrm}$\,, then the relics have less than the dark-matter density providing \cite{Carr:1994ar}
\begin{equation}
	f( M )
		<
								1
	\Rightarrow
								\beta'( M )
		<
								2 \times 10^{-28}\,\kappa^{-1}\,
								\left(
									\frac{M}{M_{\rm Pl}}
								\right)^{\!3/2}
								\label{eq:betarelic}
\end{equation}
for the mass range
\begin{equation}
	\left(
		\frac{T_{\Rrm}}{T_{\rm Pl}}
	\right)^{-2}
		<
								\frac{M}{M_{\rm Pl}}
		<
								10^{11}\,\kappa^{2 / 5}
								\, .
								\label{eq:relic}
\end{equation}
The lower mass limit arises because PBHs generated before reheating are diluted exponentially. The CMB quadrupole anisotropy implies $T_{R} < 10^{16}\,$GeV, so the lower limit exceeds $10^{6}\,M_{\rm Pl}$. The upper mass limit arises because PBHs larger than this dominate the total density when they evaporate, in which case the final cosmological photon-to-baryon ratio is determined by the baryon asymmetry associated with their emission.

The mass window is illustrated in Fig.~\ref{fig:windows}D for parameters similar to those used in Ref.~\cite{Carr:1994ar}. Note that $M$ here refers to the {\it initial} PBH mass and not the relic mass. It should be stressed that limit \eqref{eq:betarelic} applies even if there is no inflationary period but it then extends all the way down to the Planck mass. Since Planck-mass relics are not the usual type of black hole, this scenario is fundamentally different from the other three. At present there are no constraints for Planck-mass relics and it is hard to conceive of any in the future, since the relics are so small. Indeed, since they are the smallest conceivable objects in nature, they could never be detected non-gravitationally unless perhaps one invokes TeV quantum gravity. So this scenario is completely open and one cannot predict the mass function on the basis of the arguments used in Sec.~\ref{sec:RealisticMassFunctions}.

\section{LIGO gravitational-wave limits}
\label{sec:LIGO}

\noindent
A population of massive PBHs would be expected to generate a background of gravitational waves \cite{1980A&A....89....6C}. This would be especially interesting if there were a population of binary black holes, coalescing at the present epoch due to gravitational-radiation losses. This was first discussed by Bond and Carr \cite{1984MNRAS.207..585B} in the context of Population III black holes and later in Refs \cite{Nakamura:1997sm, Ioka:1998gf} in the context of PBHs. However, the precise formation epoch of the holes is not crucial since the coalescence occurs much later. In either case, the black holes would be expected to cluster inside galactic halos (along with other forms of dark matter) and so the detection of the gravitational waves would provide a unique probe of the halo distribution \cite{Inoue:2003di}. The LIGO data had already placed weak constraints on such scenarios a decade ago \cite{Abbott:2006zx}.

The suggestion that the dark matter could comprise PBHs in the IMBH range has attracted much attention recently as a result of the LIGO detections \cite{Abbott:2016blz, Abbott:2016nmj} of merging binary black holes with mass around $30\,M_{\odot}$. Using slightly different approaches, Refs.~\cite{Bird:2016dcv} and \cite{Clesse:2016vqa} derive merger rates for particular PBH populations and find them to be compatible with the range $9$ -- $240\.{\rm Gpc}^{-3}\.{\rm y}^{-1}$ obtained by the LIGO analysis. However, according to Ref.~\cite{Eroshenko:2016hmn}, the PBH merger rates would be highly suppressed by tidal forces, so that the LIGO results allow only a small fraction of the dark matter to be in PBHs. This conclusion is also drawn in Ref.~\cite{Sasaki:2016jop}, which points out that the lower limit on the merger rate may be in tension with the CMB distortion constraints \cite{Ricotti:2007au} for objects in the IMBH range. This could exclude PBHs as sources of the observed mergers. However, the accretion and merger of smaller PBHs after decoupling might still provide a PBH population like GW150914 without violating the CMB constraints \cite{Clesse:2016vqa}.

Ref.~\cite{Raccanelli:2016cud} suggests a scheme for distinguishing between black-hole mergers of stellar and primordial origin, which involves matching their spatial distribution with galaxy catalogue data. However, this could be implemented only if future merger events are more precisely localised and ascertaining the location and mass distribution of LIGO events will be difficult \cite{Dai:2016igl}. Although it is unclear whether the black holes associated with the LIGO events are of primordial or stellar origin, it is important to stress that the constraints are now based on a clear detection rather than a null result. Future LIGO and other gravitational-wave detector data will provide improved constraints.

Recently, the prospect of eLISA detecting gravitational waves from binary black holes like GW150914 in the Galaxy has been discussed by Seto \cite{Seto:2016wom}, including a potential measurement of the eccentricity down to $e \approx 0.02$. Nishizawa {\it et al.} also discuss how eLISA eccentricity measurements can constrain stellar binary black-hole formation scenarios \cite{Atsushi:160607631}. Kyutoku and Seto \cite{Kyutoku:2016ppx} claim that eLISA might observe as many GW150914-type binary black holes as supermassive binary black holes, although most of them will not merge within the eLISA observation period. Apart from eLISA, other future space missions, such as the Japanese space gravitational-wave antennas (Pre-)DECIGO (DECi hertz laser Interferometer Gravitational-wave Observatory) \cite{Kawamura:2006up} might be able to distinguish between binary black holes of Population II, Population III or primordial origin \cite{Nakamura:2016hna}. This is because (Pre-)DECIGO will be able to measure the mass spectrum and $z$-dependence of the merger rate. For example, $30\,M_{\odot}$ binary mergers like GW150914 will be detected up to redshifts $z \approx 30$ and it may be able to localise the direction of the binary black holes at $z = 0.1$ with an accuracy of $0.3\,{\rm deg}^{2}$ \cite{Nakamura:2016hna}.

Stochastic gravitational-wave backgrounds from black-hole binaries offer another way of distinguishing between the progenitors of binary black-hole mergers \cite{TheLIGOScientific:2016pea} observed by advanced LIGO. However, the PBH profile has yet to be worked out and the information needed may be difficult to extract \cite{Callister:2016ewt}. For an updated analysis of the presence of such a stochastic background in the light of the GW150914 and GW151226 merger events, see Refs.~\cite{TheLIGOScientific:2016pea} and \cite{TheLIGOScientific:2016wyq}. Finally, since PBHs probably do not {\it form} in binaries, their orbital eccentricities might make their gravitational-wave merger signal distinguishable from that of astrophysical black-hole binaries \cite{Cholis:2016kqi}.

A different type of gravitational-wave constraint on $f( M )$ has been pointed out by Saito and Yokoyama \cite{Saito:2008jc}. This is because the second-order tensor perturbations generated by the scalar perturbations which produce the PBHs are surprisingly large. The associated frequency was originally given as $10^{-8}\,( M / 10^{3}\,M_{\odot})\,{\rm Hz}$ but this estimate contained a numerical error \cite{Saito:2009jt} and was later reduced by a factor of $10^{3}$ \cite{Bugaev:2009zh}. The limit on $f( M )$ just relates to the amplitude of the density fluctuations at the horizon epoch which is of order $10^{-52}$. This effect has subsequently been studied in Ref.~\cite{Assadullahi:2009jc} and by several other authors. In particular, the limit from pulsar timing data already excludes PBHs with $0.03\,M_{\odot} < M < 10\.M_{\odot}$ from providing an appreciable amount of dark matter \cite{Bugaev:2010bb} and limits from LIGO, VIRGO and BBO could potentially cover the mass range down to $10^{20}\,\grm$. Conversely, one can use PBH limits to constrain a background of primordial gravitational waves \cite{Nakama:2015nea, Nakama:2016enz, Pen:2015qta}.

None of these limits is not shown in Fig.~\ref{fig:large} because they apply only if the PBHs are generated by super-Hubble scale fluctuations, such as arise through inflation. However, this is the most popular scenario for PBH formation, which is why these limits were included in Fig.~8 of Carr \textit{et al.} \cite{Carr:2009jm}. We also note that the limiting value of $f$ depends on the fluctuations being Gaussian. Although this is questionable in the context of the large-amplitude fluctuations relevant to PBH formation, the studies in Refs.~\cite{Hidalgo:2007vk, Hidalgo:2009fp} show that non-Gaussian effects are not expected to be large.

\section{Summary \& Outlook}
\label{sec:Summary-and-Outlook}

\noindent
In this work we have studied the possibility that PBHs constitute the dark matter, focussing on the three mass ranges where PBHs were considered plausible dark-matter candidates around a decade ago. These include (A) black holes in the intermediate-mass range $1\,M_{\odot} < M < 10^{3}\,M_{\odot}$, (B) sublunar black holes in the range $10^{20}$ -- $10^{24}\,$g and (C) sub-atomic size black holes in the range $10^{16}$ -- $10^{17}\,$g. In addition, we have discussed (D) Planck-mass relics in the range around $10^{-5}\,$g. All relevant constraints in these mass windows were reviewed in Sec.~\ref{sec:Constraints}, including those from microlensing, dynamical effects, large-scale structure, accretion and black-hole mergers of the kind observed by LIGO. We have found that scenarios (A) and (B) can still produce all the dark matter, although this depends on the exact values of the astrophysical parameters involved in the constraints. So these windows may be closed in the near future. Scenario (C) is already excluded for all practical purposes, while (D) is completely unconstrained and will remain so for the foreseeable future.

Since the precision of the constraints has improved significantly in recent years, a more refined treatment of PBH formation appears to be mandatory. In order to tackle this issue, we discussed in Sec.~\ref{sec:RealisticMassFunctions} all the necessary ingredients for a precise calculation of the PBH abundance from a fundamental early-Universe source, such as non-Gaussianity, non-sphericity, criticality, merging, the choice of the appropriate variables, and the different approaches for estimating the black-hole number density. Regarding non-Gaussianity, non-sphericity and criticality, we have performed quantitative calculations, showing how these effects are expected to change the PBH distribution. In all cases the mass spectrum will be lowered, while critical collapse will cause significant broadening, as well as a shift towards lower masses.

In Sec.~\ref{sec:ExtendedPBH} we introduced a novel scheme for investigating the compatibility of a general extended PBH mass function with arbitrary constraints. We also showed which model-independent conclusions can be drawn from the constraints for an unknown extended mass function, illustrating this by the application to constraints in the intermediate-mass region. Our procedure demonstrated, on the one hand, that extended mass spectra are more difficult to analyse than the commonly (and wrongly) used monochromatic ones. On the other hand, we showed that there are situations in which PBH dark matter is excluded in the monochromatic case but allowed in the extended mass case. We have given explicit examples of this. For definiteness, we introduced in Sec.~\ref{sec:ModelsInRelevantRanges} two inflationary models{\,---\,}the axion-like curvaton model and the running-mass model){\,---\,}which are capable of producing PBHs in the relevant mass ranges. In Sec.~\ref{sec:ExtendedPBH} we confronted these models with the latest constraints in these mass ranges and discussed under what circumstances they can produce PBHs containing all the dark matter.

Even though we have presented a rather complete picture of PBH formation, more work is required for the concrete implementation of this approach. In particular, for non-sphericity, precision simulations of fully general-relativistic collapses for ellipsoidal overdensities are necessary. Additional clarification of the interplay between ellipticity and non-Gaussianity, as well as a more thorough understanding of merger rates and accretion are needed before the observational constraints on PBHs can be translated into constraints on early-Universe physics.

Even before all issues of PBH formation are settled, model-independent exclusion of PBHs as dark-matter candidates may be possible in the near future. If care is taken when applying observational constraints to allow for uncertainties in the various astrophysical processes (\eg~the growth of the PBH through accretion), then one may be able to exclude even PBHs with extended mass functions. However, this must be done by considering constraints in the way described in this paper, rather than by focusing on monochromatic mass functions which contain all the dark matter.

\acknowledgments

We thank Chris Byrnes, Paul Frampton, Anne Green, Phillip Helbig, Alex Kusenko, John Miller, Paul Schechter and Yuichiro Tada for useful discussions. The review of the constraints in Sec.~V updates work originally presented in Ref.~\cite{Carr:2009jm}, so we thank Kazunori Kohri, Yuuiti Sendouda and Jun'ichi Yokoyama for helpful input. F.K.~acknowledges support from the Swedish Research Council (VR) through the Oskar Klein Centre.

\bibliography{refs}

\end{document}